\documentclass[twocolumn]{aastex61}

\newcommand{\tic}{$T_\mathrm{IC}$}

\newcommand{\srcname}{H~1743$-$322}

\newcommand{\ledd}{$L_\mathrm{Edd}$}
\newcommand{\nh}{$N_\mathrm{H}$}
\newcommand{\rxte}{{\it RXTE}}
\newcommand{\swift}{{\it Swift}}
\newcommand{\chandra}{{\it Chandra}}
\shorttitle{Disk winds in H J1743$+$322}
\shortauthors{Shidatsu et al.}

\begin{document}

\title{Application of the Thermal Wind Model to Absorption Features in the Black Hole X-ray Binary H~1743$-$322}

\correspondingauthor{Megumi Shidatsu}
\email{megumi.shidatsu@riken.jp}

\author{Megumi Shidatsu}
\affiliation{Department of Physics, Ehime University, Matsuyama 790-8577, Japan}
\author{Chris Done}
\affiliation{Department of Physics, University of Durham, South Road, Durham, DH1 3LE, UK}



\begin{abstract}
High inclination black hole X-ray binaries exhibit
blueshifted ionized absorption lines from disk winds, 
whose launching mechanism is still in debate.
The lines are predominantly observed in the high/soft state 
and disappear in the low/hard state, anti-correlated with the jet. 
We have tested if the thermal winds, which are driven by 
the irradiation of the outer disk by the X-rays from 
the inner disk, can explain these observed properties
or whether we need a magnetic switch between jet and wind. We use analytic thermal-radiative wind models to 
predict the column density, ionisation parameter and velocity of the wind given the 
broadband continuum shape and luminosity determined from RXTE monitoring. We use these to simulate the detailed photo-ionised absorption features predicted at epochs where there are  {\it Chandra} high resolution spectra. These include low/hard, high/soft and very high states. 
The model was found to well reproduce the observed lines in the high/soft state, and also successfully 
predicts their disappearance in the low/hard state. However, the simplest version of the thermal wind model also predicts that there should be strong features observed in the very high state, which are not seen in the data. Nonetheless, we show this is consistent with thermal winds when we include self-shielding by the irradiated inner disk atmosphere.
These results indicate that the evolution of observed wind properties in different states 
during outbursts in H 1743$-$322 can be explained by the thermal wind model and does not require magnetic driving.
\end{abstract}


\section{Introduction} \label{sec:intro}

Disk winds have been observed in several 
black hole X-ray binaries (BHXBs) as 
blue-shifted, highly ionized absorption 
lines, especially H- or He-like iron-K 
lines, on the X-ray continuum spectra 
\citep[e.g.,][]{kot00, ued01, mil06, kub07, mil08, 
dia14, hor18}. They are only seen in high 
inclination systems, suggesting that the 
winds have an equatorial structure, extending 
along the disk plane with a small solid angle 
\citep{pon12}. The observed winds have state 
dependence; the absorption lines are predominantly 
seen in the high/soft state, and they tend to be 
more ionized with spectral hardening \citep{dia14,hor18} 
and finally disappear in the low/hard 
state \citep{mil08,nei09,pon12,mil12}. 

What drives the winds in BHXBs is 
a long-standing question. 
Radiation pressure by Compton scattering can 
drive winds when it overcomes the gravity 
of the central black hole. This mechanism, 
by definition, only works above the Eddington 
luminosity ($L_\mathrm{Edd}$) but most of 
the systems are well below $L_\mathrm{Edd}$ hence 
this continuum-radiation-pressure 
driven wind is unlikely to explain  
the majority of the disk winds seen in BHXBs. 
Radiation pressure on bound-free or line transitions can 
launch a wind below $L_\mathrm{Edd}$, giving a plausible
mechanism for some winds 
in active galactic nuclei \citep{pro00, nom16}, 
but it is again unlikely to work in BHXBs 
because their much higher temperature disks
mean that the strong UV absorption species are
completely ionized \citep{pro02}.

Instead, a promising launching mechanism of winds in BHXBs  
is thermal driving. The outer disk regions are  
irradiated by the strong X-rays emitted from 
the inner disk region. Gas in the disk photosphere 
is then heated to the Compton temperature $T_\mathrm{IC}$, 
where Compton up and down scattering is balanced. This 
temperature is determined by the shape of the 
spectral energy distribution, as 
\begin{equation}
   T_\mathrm{IC} =  \frac{ \int_0^\infty h\nu L_\nu d\nu}{4 k \int_0^\infty L_\nu d\nu},
\end{equation} 
where $h$ is the Planck constant and 
$k$ is the Boltzmann constant \citep[see e.g.,][]{beg83,don10}.
Its typical value for BHXBs is $\sim 10^7$ K 
in the high/soft state. This gas can escape from 
the disk when its kinetic energy overcomes the 
local gravitational energy. This gives 
an estimate for the wind launching radius $R_\mathrm{IC} 
\sim G M_\mathrm{BH} \mu/k T_\mathrm{IC}$, 
where $\mu\sim 0.6m_p$ is the mean particle mass 
in the wind \citep{beg83, woo96}.
If the outer disk radius $R_\mathrm{out}$ 
is smaller than $R_\mathrm{IC}$, the illuminated 
gas kept bound on the disks forming a 
static ionized atmosphere above the disks, as indeed 
observed in many short period (mainly neutron star) 
X-ray binaries \citep{dia13}, 
whereas winds are observed only in systems with big disks \citep{dia16}.

The final mechanism, magnetic driving, has 
drawn growing interest since the discovery of a peculiar wind 
in GRO J1655$-$40 
\citep{mil06, mil08, fuk17}, in which the 
wind launching radius calculated from the 
absorption features was much smaller than 
$R_\mathrm{IC}$. This idea, that the magnetic 
fields powers the winds, also led attempts to 
explain the observed state dependence 
of the wind properties as an anti-correlation 
with the jet, so that the same magnetic field 
reconfigures to power the jet in the low/hard 
state and the wind in the high/soft state \citep{fuk14}.
However, their launching 
site is very different; winds are generally launched 
in the outer disk regions, whereas jets are 
believed to be powered in the innermost 
regions of the disk, and hence it is not likely that 
they are really associated via the 
same magnetic fields. Also, recent studies 
suggests that the peculiar wind in 
GRO J1655$-$40 may be explained by a 
Compton-thick, thermal (plus 
continuum-radiation-pressure) driven wind 
\citep{utt15, nei16, shi16}. 

\begin{figure*}[htb]
\epsscale{1.2}
\plotone{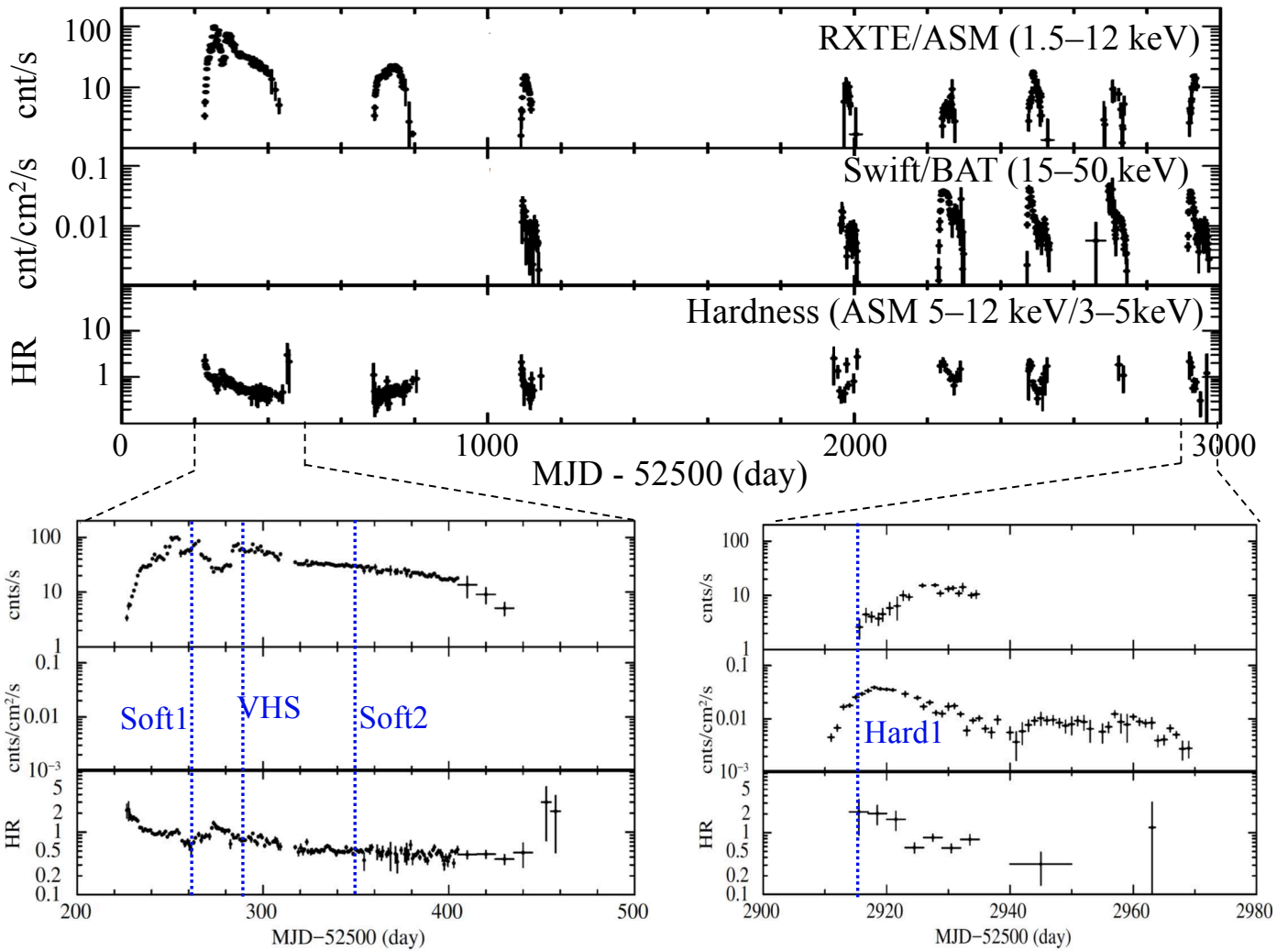}
\caption{Long-term light curves and hardness ratio of H 1743$-$322 until 2011. The bottom panels are enlarged views of the top panel. 
The \chandra\ observations before 2011 considered in this work 
are indicated in blue dotted lines. 
At Hard2 in 2015, the \swift/BAT count rate was $\sim 0.01$ counts cm$^{-2}$ s$^{-1}$.
\label{fig:lc_longterm}}
\end{figure*}

Given that the thermal winds are relatively well 
understood theoretically \citep{beg83, woo96, hig14}, 
compared with the magnetic winds, one possible approach 
would be to study to what extent the thermal winds 
can describe the observed absorption features 
and its state dependence, and then explore 
how much room remains to invoke magnetic winds. 
\citet{don18} (hereafter D18) provided a predictive 
thermal wind model, which can derive the basic wind 
parameters including the column density and the ionization 
parameter. They set up a simplified spectral model,
where the continuum depends only on $L/L_\mathrm{Edd}$, 
such that it was dominated by a disk with $L\propto T^4$ in the
high/soft state, switching to a power law for the low/hard state 
at $L/L_\mathrm{Edd}=0.02$. They concluded that the 
resultant thermal (and thermal-radiative) 
wind properties could explain 
most (and perhaps all) of the currently available data.
However, the actual spectral evolution in BHXBs 
is more complex, and not determined by luminosity alone. 
The high luminosity states are not always dominated by the 
disk emission as assumed in D18, but can have a more 
substantial soft Compton tail (very high state). Also, the 
transition to the low/hard state is not at a fixed luminosity, 
as displayed by the hysteresis seen in the hardness 
intensity diagram.
These different SED-Luminosity behaviours will change 
the predicted wind properties since 
the thermal winds are very sensitive to the shape of the 
continuum spectrum as well as its luminosity. 

Here, we instead use the actual X-ray data 
of the BHXB H 1743$-$322 taken in monitoring 
observations with \rxte\ and \swift,
to accurately determine the continuum spectral shape 
and luminosity throughout the outbursts. 
We then predict the thermal wind parameters 
(column density, ionization state and velocity) 
appropriate for each spectrum using the D18 model, to  
predict how the thermal winds evolve across a real outburst. 
There are also several 
\chandra\ high resolution spectra taken in different states, 
including the high/soft state where the wind features were visible, 
and the low/hard and very high states where they were not 
significantly detected.
We use photo-ionization models to compare the detailed 
predictions of the thermal wind model to the high resolution 
spectra, and find that they are a good match to the 
observations. We conclude that these winds are most likely 
thermally driven rather than powered by magnetic fields. 

\section{System Parameters and Long-term X-ray Properties of H 1743$-$322} \label{sec:1743}

We first summarize the X-ray and binary system properties of 
H 1743$-$322. This is one of the systems in which
winds have been detected (see e.g., \citealt{pon12}).

This source has exhibited many outbursts which have been 
extensively observed at various wavelengths especially with 
{\it Rossi X-ray Timing Explorer} (\rxte) and 
\swift. These also provide daily broad-band X-ray monitoring 
data covering the entire outburst periods. 
Figure~\ref{fig:lc_longterm} presents X-ray light curves 
in 1.5--12 keV from the \rxte/All Sky Monitor 
(ASM) and in 15--50 keV from the \swift/Burst Alert Telescope (BAT). 
The ASM hardness ratio (HR) between 5--12 keV and 3--5 keV is 
also shown as the lower panel in Fig.~\ref{fig:lc_longterm},
with state transitions to the high/soft state indicated by $HR\lesssim 1$.

\begin{deluxetable}{ccccc}[ht!]
\tablecaption{List of \chandra/HETGS observations considered in this work \label{tab:obslist_chandra}}
\tablecolumns{5}
\tablewidth{0pt}
\tablehead{
\colhead{Epoch} & \colhead{OBSID} & \colhead{Date} & \colhead{State} & \colhead{Lines?} 
}
\startdata
Soft1 & 3803 & 2003 May 1--2 & high/soft & yes \\
VHS & 3804 & 2003 May 28 & very high & no \\
Soft2 &  3806 & 2003 Jul. 30--31 & high/soft & yes \\
Hard1 & 11048 & 2010 Aug. 8--9 & low/hard & no \\ 
Hard2\tablenotemark{a} & 16738 & 2015 Jul. 11 & low/hard & no \\
Hard2\tablenotemark{a} & 17679 & 2015 Jul. 12 & low/hard & no \\
Hard2\tablenotemark{a} & 17680 & 2015 Jul. 13 & low/hard & no \\
\enddata
\tablenotetext{a}{The luminosities and SED profiles are almost the same in the three observations and their spectra are co-added to improve statistics in Section~\ref{sec:simulation}.}
\end{deluxetable}

\chandra\ carried out high resolution spectroscopy 
several times in these outbursts as listed in Table~\ref{tab:obslist_chandra}. These sample different spectral state
and luminosities. Observations with 
too low statistics are omitted. As given in this table, 
we hereafter call these \chandra\ epochs Soft1, Soft2, 
VHS, Hard1, and Hard2. 
The three sequential observations in 2015 (Hard2), 
where the luminosity and SED profile do not differ 
significantly, are combined to obtain high resolution 
spectrum in the low/hard state at a relatively low 
luminosity, although \rxte\ already ended its 
operation and the broad-band continuum data are unavailable.
Hence we estimate the HR and luminosity by matching to 
RXTE data at similar 3-7~keV continuum shape and luminosity. 
The H-like and He-like Fe K$\alpha$ absorption lines 
are clearly detected only in the high/soft state (Soft1 and 2), 
whereas no significant lines were detected in the 
other epochs \citep{mil06b,mil12}. 

The inclination angle and the distance of 
H 1743$-$322 were constrained by \citet{ste12} 
from the trajectory of ballistic jets as 
$75^\circ \pm 3^\circ$ and $8.5 \pm 0.8$ kpc, 
respectively. The high inclination angle is supported 
by the fact that the source shows absorption dips 
in its X-ray light curves and ionized 
absorption lines from winds in its spectra. 
Short-term variability properties also imply 
a high inclination angle; the source shows a somewhat 
stronger low frequency QPOs in the low/hard state 
than low inclination BHXBs, as expected if the QPO is a 
geometric effect such as Lense-Thirring precession 
\citep{ing09}. \citet{ste12} estimated
the black hole mass as $\sim 7 M_\sun$ from disk continuum 
fits with a relativistic accretion disk emission model 
(assuming spin parameter, $a_* = 0.2$). 

The outer disk radius $R_\mathrm{out}$, is poorly known, but this is a 
critical parameter for calculation of the thermal winds. We estimate 
this from comparison of the frequency of outbursts to disk instability 
calculations. The multiple outbursts suggest that the mass transfer 
rate from the companion star is close to the critical mass accretion rate 
where the hydrogen ionization instability is triggered \citep{cor12}. 
GX339$-$4 is similarly a system which 
shows frequent outbursts, so we assume that the 
orbital period of H 1743$-$322 
is similar to that of GX 339$-$4 ($\sim$40 hours).
Thus the disk would similarly extend to 
a few tens of percent of its Roche lobe, giving an estimate for
$R_\mathrm{out}\sim 3 \times 10^{11}$ cm, 
but this must be uncertain by at least a factor 2 in either 
direction. This disk size is likely the smallest among 
the BHXBs in which winds have been detected (\citealt{pon12}), 
and hence it is the simplest to model (see also \citealt{tom19}a, b).

We note that D18 used different system parameters for this source, 
with a  black hole mass of $10\ M_\odot$ and spin of $0.5$ at a distance 
of $5$~kpc. Most importantly, they assumed $R_\mathrm{out}=3.7 \times 10^{12}$~cm, 
almost an order of magnitude larger than here. The predicted column 
density in the wind material is $\propto \log R_\mathrm{out}/R_\mathrm{wind}$ 
(where $R_\mathrm{wind}$ is the wind launching radius), 
so typically our columns will be a factor $\sim 1.8 \times$ smaller 
for a given $L/L_\mathrm{Edd}$.

\section{Modeling Continuum X-ray Spectra}\label{sec:xrtana}

We produced broadband X-ray spectra corresponding to each 
pointed {\rxte}/PCA observation of \srcname. These were 
extended to higher energies using {\rxte}/HEXTE (up to 
2010) or {\swift}/BAT (after 2010). 

\begin{figure*}[htb]
\epsscale{1.1}
\plottwo{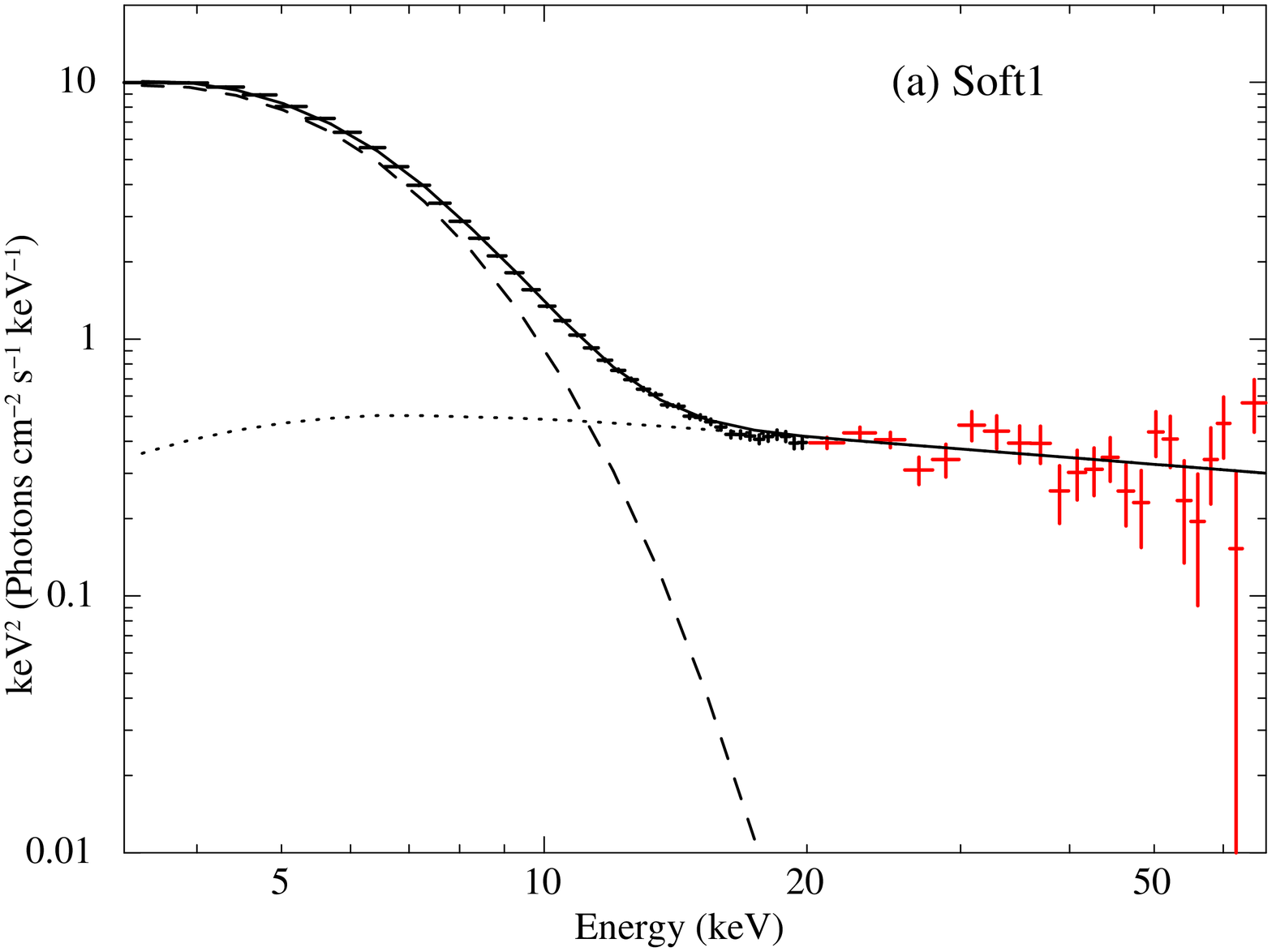}{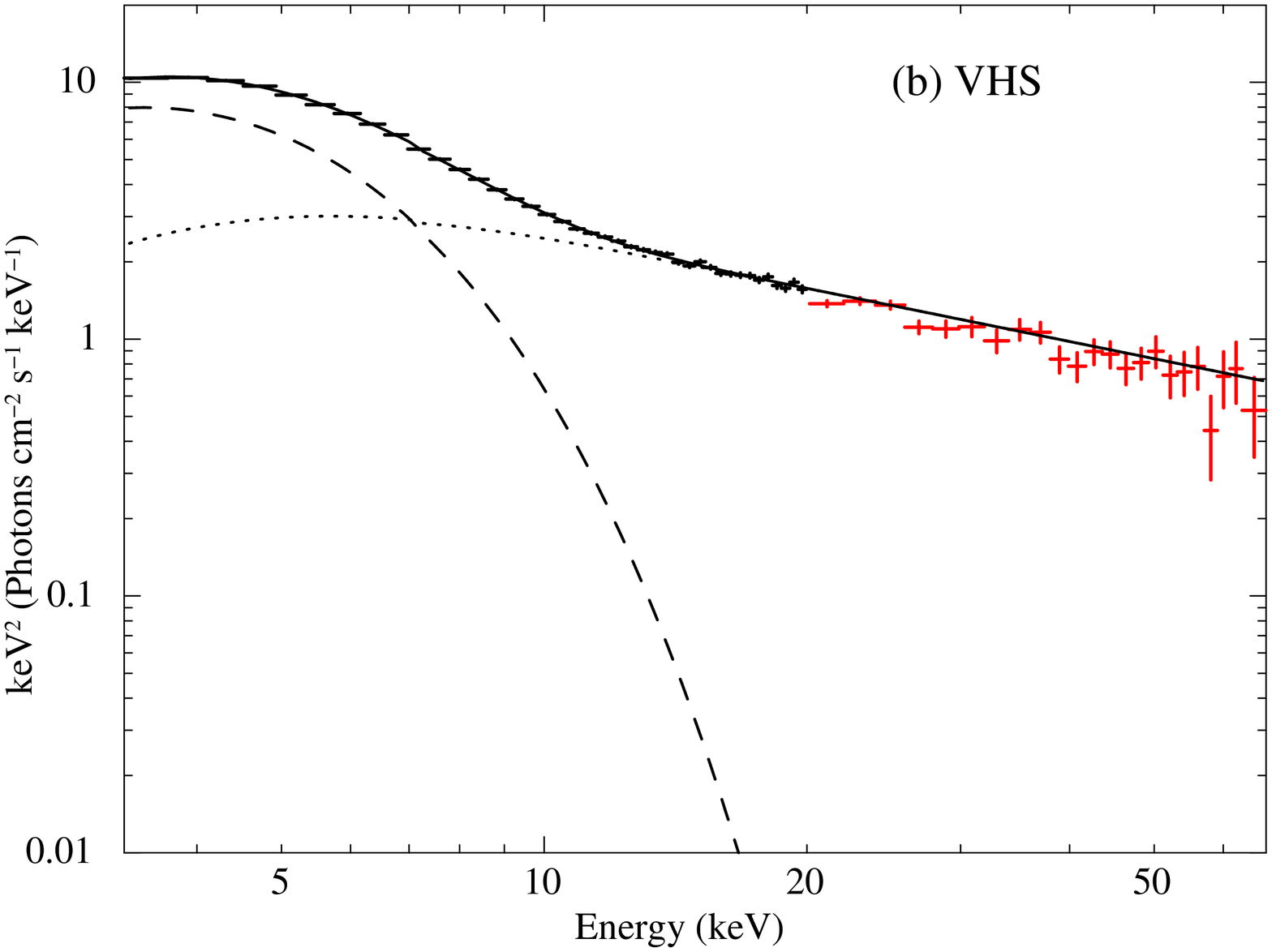}
\plottwo{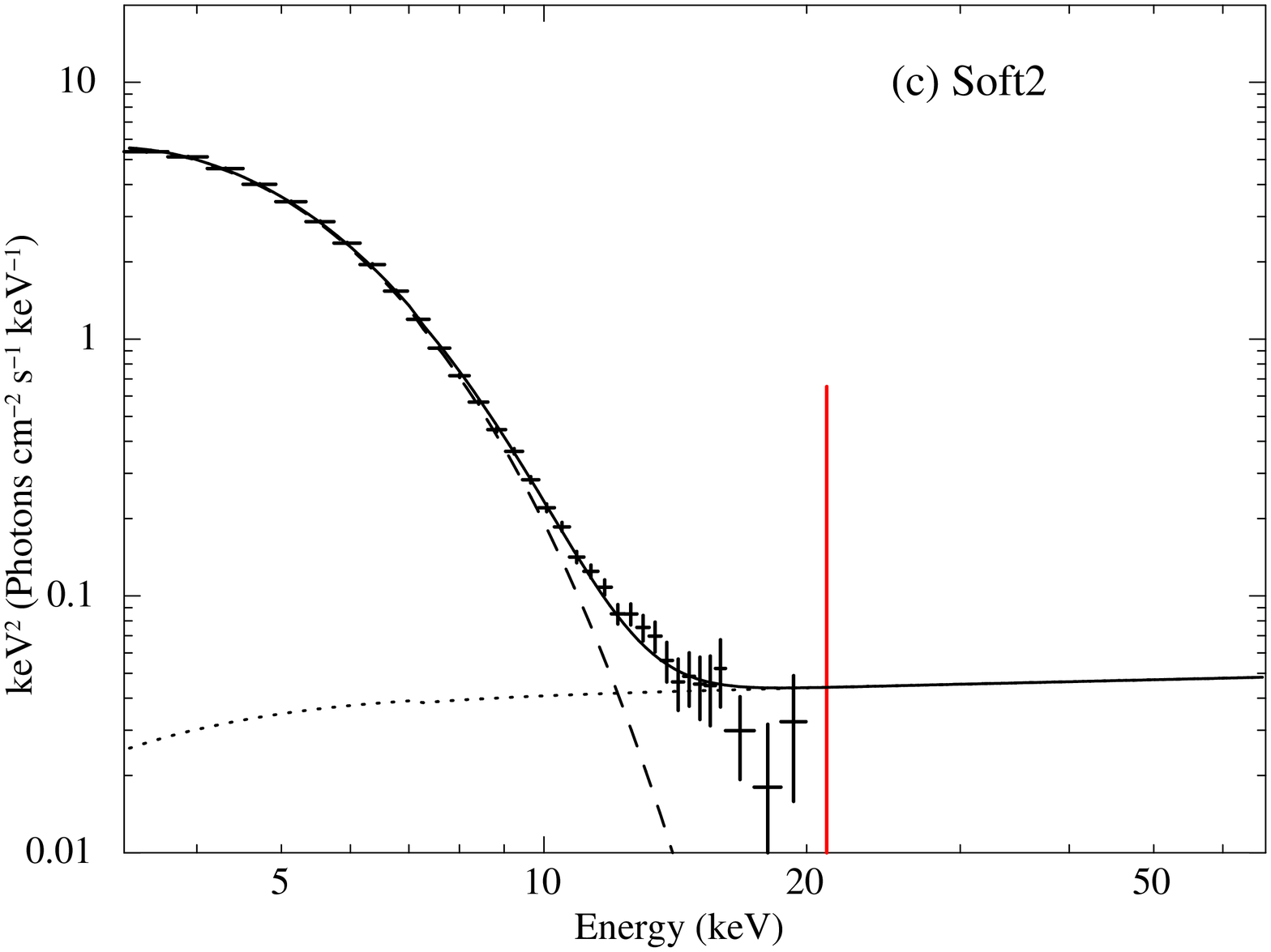}{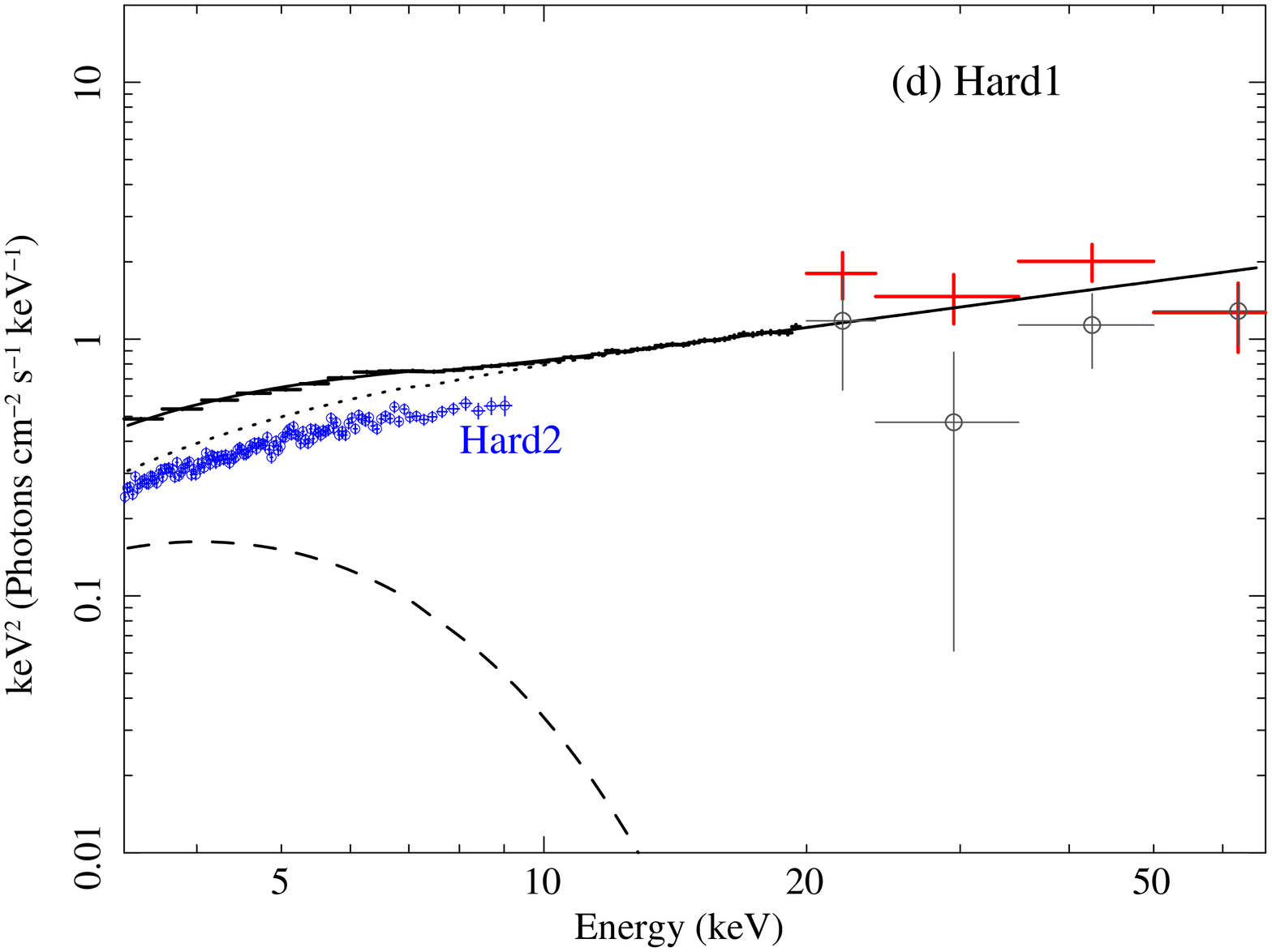}
\caption{The broadband continuum spectra of H 1743$-$322 
at the four \chandra\ epochs in 2003--2010, 
with their best-fit {\tt tbabs*simpl*diskbb} models. 
The \rxte/PCA data are shown in black, and the \rxte/HEXTE 
Cluster-A (in panel a--c) and \swift/BAT data (in panel d) in red. 
The HEXTE Cluster-B data are omitted in the panel (a--c) for 
illustrative purposes. In panel (d) the \chandra\ HEG spectrum 
and a \swift/BAT spectrum at Hard2 are also plotted with blue 
and grey circles, respectively, for comparison with Hard1 (see 
Sec.~\ref{sec:simulation} for the details of the \chandra\ data). \label{fig:continuum}}
\end{figure*}

The {\rxte} data were reduced in the standard manner 
described in the {\rxte} cookbook, by using HEAsoft 
version 6.19 and the Calibration Database 
(CALDB) downloaded in 2016 December. 
We extracted the PCA spectra from the ``Standard 2'' data 
of the Proportional Counter Array 2 (PCA2) and the HEXTE 
spectra from Clusters A and B data. 
To obtain hard X-ray spectra after 2009 December, 
when the \rxte/HEXTE stopped rocking between 
the on-source and off-source positions, we used 
the \swift/BAT survey data taken on the same day 
as \rxte/PCA data. The BAT survey data were 
downloaded from the HEADAS 
archive\footnote{\url{https://heasarc.gsfc.nasa.gov/FTP/swift/data/obs/}}
and processed with the ftool {\tt batsurvey} 
referring to the latest {\swift} CALDB as of 2016 
December. The spectra and their response files 
were generated from the individual continuous scans 
using the script {\tt make\_survey\_pha}. 
We chose the scan with the longest exposure if multiple 
scans were present.
In this way, we obtained $\sim$500 simultaneous 
broad-band X-ray spectra of H 1743$-$322, 
covering 8 outbursts from 2003 March to 2011 April.

Figure~\ref{fig:continuum} presents the resulting 
broadband continuua 
corresponding to the \chandra\ high resolution datasets 
in Soft1, Soft2, VHS, and Hard1. In Fig.~\ref{fig:continuum}(d) 
we also present the \chandra\ HEG spectrum at Hard2 
and a corresponding \swift/BAT spectrum taken 
on 2015 July 12. 
The Soft1 and Soft2 spectra are both high/soft state, 
dominated by the disk blackbody component, especially 
Soft2 which has an extremely weak hard tail. 
The VHS spectrum, taken at the highest luminosity 
among the four epochs, can be approximated by a steep 
power-law model, indicating that the source was 
in the very high state. Hard1 shows a typical 
low/hard state spectrum with a hard power-law shaped profile. 
Hard2 has a slightly harder and dimmer continuum in the 3--9~keV range,  
and is characterized by a power-law model with a photon 
index of $\sim 1.33$.

We analyzed the individual broad-band X-ray continuum 
spectra in XSPEC version 12.9.0n, with a model 
consisting of the multi-color disk blackbody emission 
({\tt diskbb}: \citep{mit84}) and its
Comptonization component
({\tt simpl}: \citep{ste09}).
The {\tt simpl} Comptonisation model 
convolves a fraction of an input spectrum 
into a power-law, using the photon index ($\Gamma$) 
and the fraction of the total input X-ray flux 
that is scattered ($F_\mathrm{scat}$).
We accounted for interstellar absorption by multiplying the 
resulting {\tt simpl*diskbb} model 
by {\tt TBabs} \citep{wil00}
with fixed $N_H=1.6\times 10^{22}$~cm$^{-2}$ \citep{cap09}.
We checked that allowing this column to be free
gives consistent results, with most of the observations 
giving values within $\sim 5 \times 10^{21}$ cm$^{-2}$ and 
$\sim 3 \times 10^{22}$ cm$^{-2}$. 
However, some spectra around the state transitions gave more 
discrepant results, but these are most likely due to our 
continuum model being too simple for these complex spectra 
rather than to any  additional neutral column intrinsic to the source. 
We checked that the slight change in best fit spectral parameters 
did not affect the overall trends in wind parameters 
in Section~\ref{sec:wind_par} and the XSTAR simulation 
results in Section~\ref{sec:simulation}.

\begin{deluxetable*}{ccCCCC}[ht!]
\tablecaption{Best-fit parameters of the continuum spectra 
at the individual \chandra~epochs, and their wind parameters 
obtained with the D18 model\label{tab:pars}}
\tablecolumns{5}
\tablewidth{0pt}
\tablehead{
\multicolumn{2}{c}{Epoch} & \colhead{Soft1} & \colhead{VHS} & \colhead{Soft2} & \colhead{Hard1} \\
\multicolumn{2}{c}{State} &
\colhead{high/soft} &
\colhead{very high} & \colhead{high/soft} & \colhead{low/hard}
}
\startdata
\multicolumn{2}{l}{\bf Best-fit continuum parameters} &&&& \\
{\tt TBabs} & $N_\mathrm{H}$ ($10^{22}$ cm$^{-2}$) & 1.6~\mathrm{(fixed)} & 1.6~\mathrm{(fixed)} & 1.6~\mathrm{(fixed)} & 1.6~\mathrm{(fixed)} \\
{\tt simpl} & $F_\mathrm{scat}$ & 2.4^{+0.1}_{-0.2} \times 10^{-2} & 0.170 \pm 0.006 
& 3^{+3}_{-1} \times 10^{-3} & 0.652^{+0.020}_{-0.001} \\
& $\Gamma$ & 2.27 \pm 0.06 & 2.69 \pm 0.03 & 1.9 \pm 0.6 & 1.55 \pm 0.03 \\ 
{\tt diskbb} & $T_\mathrm{in}$ (keV) & 1.221 \pm 0.002 & 1.189 \pm 0.006 & 1.026^{+0.004}_{-0.005} & 1.49^{+0.02}_{-0.03} \\
& norm & (8.17 \pm 0.08) \times 10^{2} & (8.8 \pm 0.2) \times 10^{2} 
& (9.9 \pm 0.2) \times 10^{2} & 15.95 \pm 0.03 \\ \hline
& $L$ ($10^{38}$ erg s$^{-1}$)\tablenotemark{a} & 3.6 & 4.7 & 2.1 & 0.67 \\
& $L/L_\mathrm{Edd}$\tablenotemark{b} & 0.33 & 0.39 & 0.20 & 0.062 \\
& $L_0$ ($10^{38}$ erg s$^{-1}$)\tablenotemark{c} & 3.5 & 3.7 & 2.1 & 0.2 \\
& \tic~($10^{8}$ K) & 0.11 & 0.16 & 0.07 & 1.0 \\ \hline 
\multicolumn{2}{l}{\bf Wind parameters} & & & & \\
& $n_0$ ($10^{12}$ cm$^{-3}$) & 1.0 & 2.9 & 0.1 & 2.1 \\ 
& $R_\mathrm{wind}$ ($10^{10}$ cm) & 6.6 & 4.0 & 13 & 1.3 \\
& $N_\mathrm{H}$ ($10^{22}$ cm$^{-2}$) & 6.7 & 12 & 1.5 & 2.6 \\
& $\xi$ ($10^4$ erg cm s$^{-1}$) & 8.1 & 8.5 & 10 & 21 \\
& $v_\mathrm{wind}$ ($10^2$ km s$^{-1}$) & 3.9 & 4.7 & 3.1 & 11 \\
\enddata
\tablenotetext{a}{Unabsorbed 0.01--100 keV luminosity, assuming a distance of 8.5 kpc.}
\tablenotetext{b}{A black hole mass of 7 $M_\sun$ is assumed (i.e., $1 L_\mathrm{Edd} = 1.1 \times 10^{39}$ ergs s$^{-1}$).}
\tablenotetext{c}{Unabsorbed 0.0136--13.6 keV luminosity, which is used in XSTAR simulations.}
\end{deluxetable*}

We extend the energy range used to calculate the model 
in XSPEC to 0.1--500 keV, 
to avoid systematic errors in the {\tt simpl} 
convolution at the upper/lower energy edges 
of the data. We discarded the data with 
3--10 keV unabsorbed fluxes below $9 \times 
10^{-11}$ erg cm$^{-2}$ s$^{-1}$ (which corresponds 
to the Eddington ratio in 0.01--100 keV of 
$L/L_\mathrm{Edd} \sim 0.002$) because the Galactic 
ridge emission was found to contaminate strongly and its 
iron K emission lines are clearly seen in the PCA 
spectra. The remaining 435 spectra was used in the 
following analysis. 

The continuum spectra are well reproduced with this 
model and we used the resulting SED from 0.01--100 keV
to calculate the Compton temperature, $T_\mathrm{IC}$, 
for each individual observation.
Figure~\ref{fig:continuum} shows these model fits 
for the broadband continuum at the 4 \chandra\ epochs 
before 2011, with the individual components shown 
separately. Model parameters and 
\tic\ are shown in Table~\ref{tab:pars}.

\begin{figure*}[htb]
\epsscale{1.1}
\plottwo{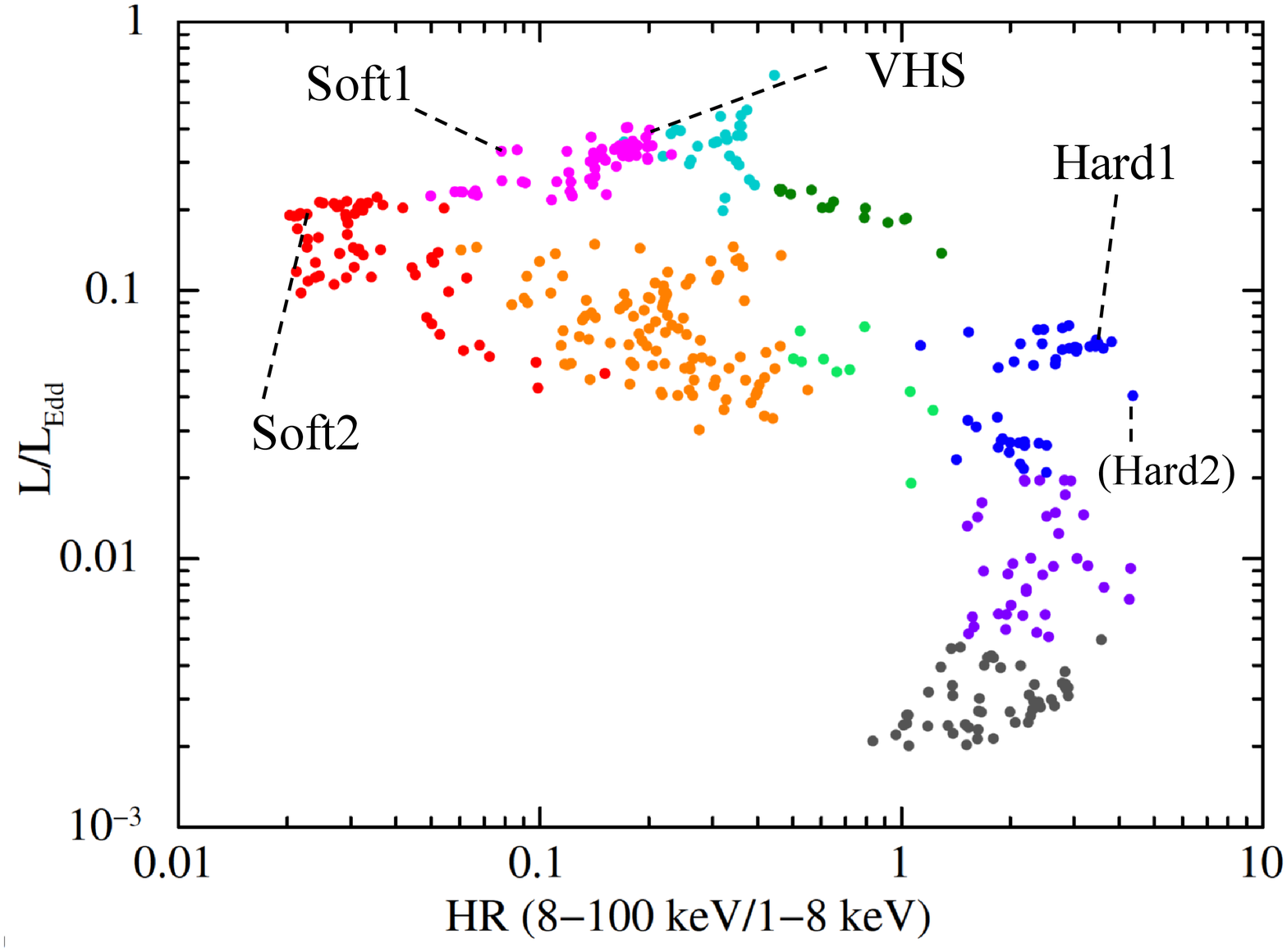}{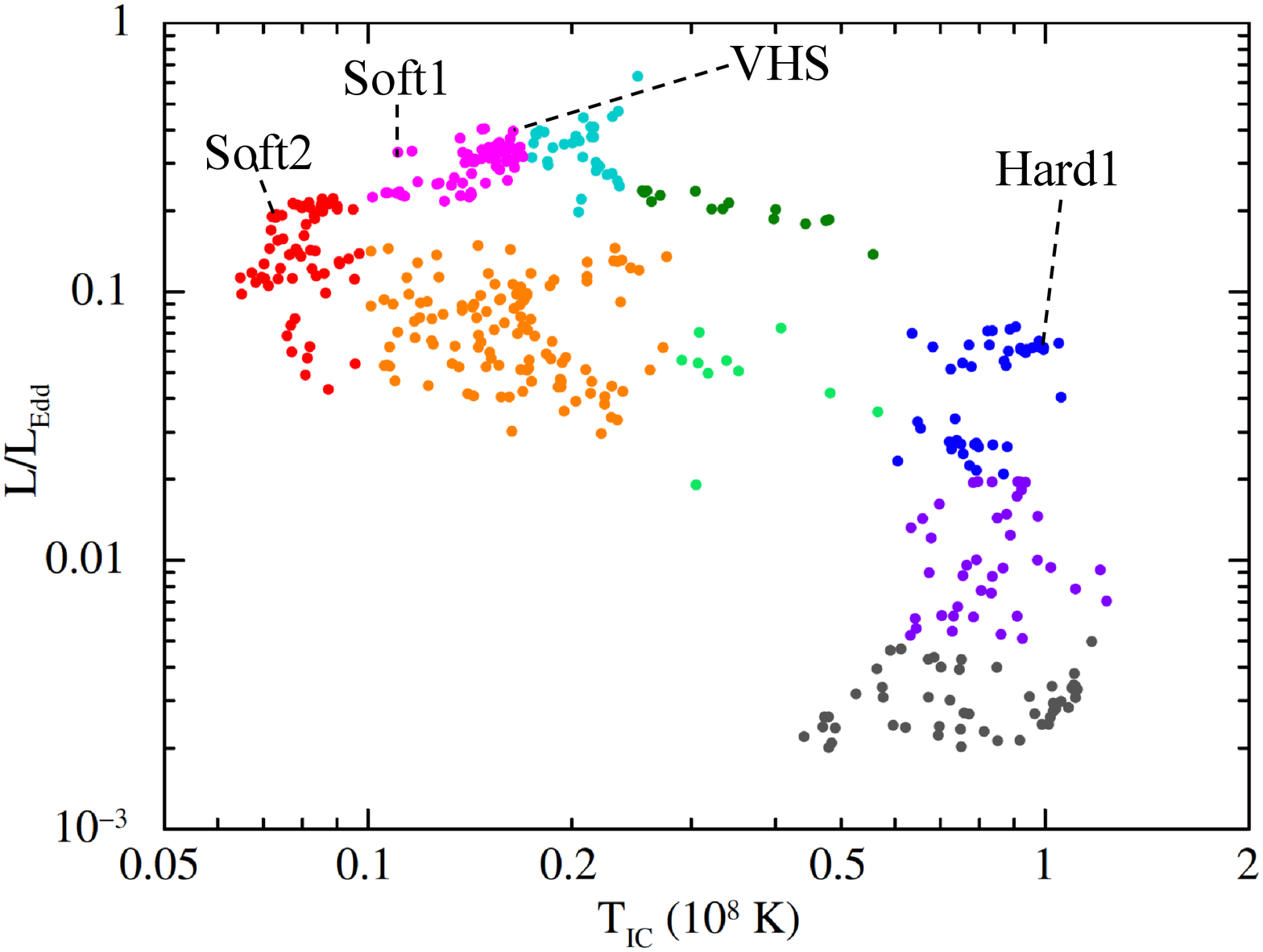}
\plottwo{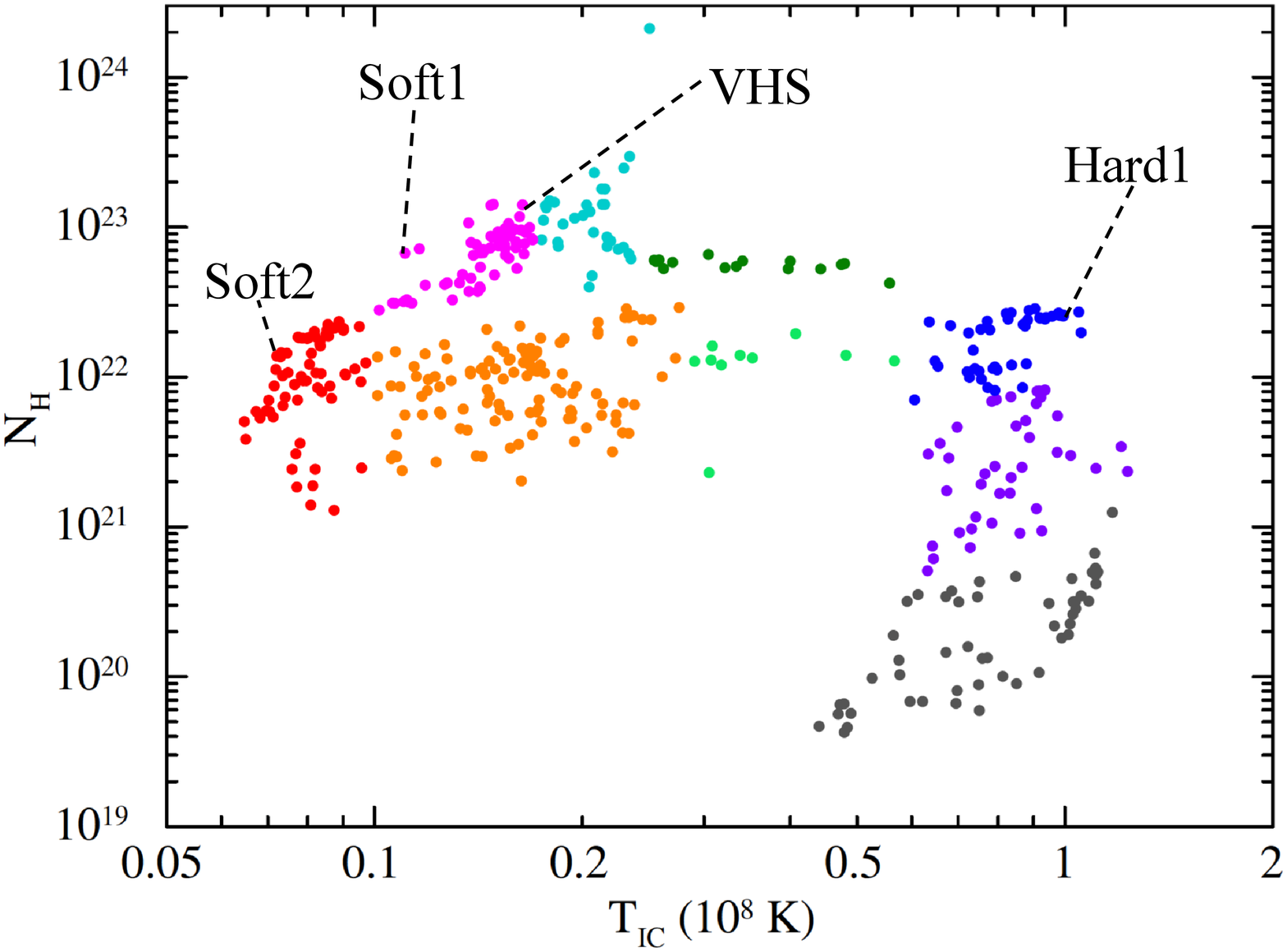}{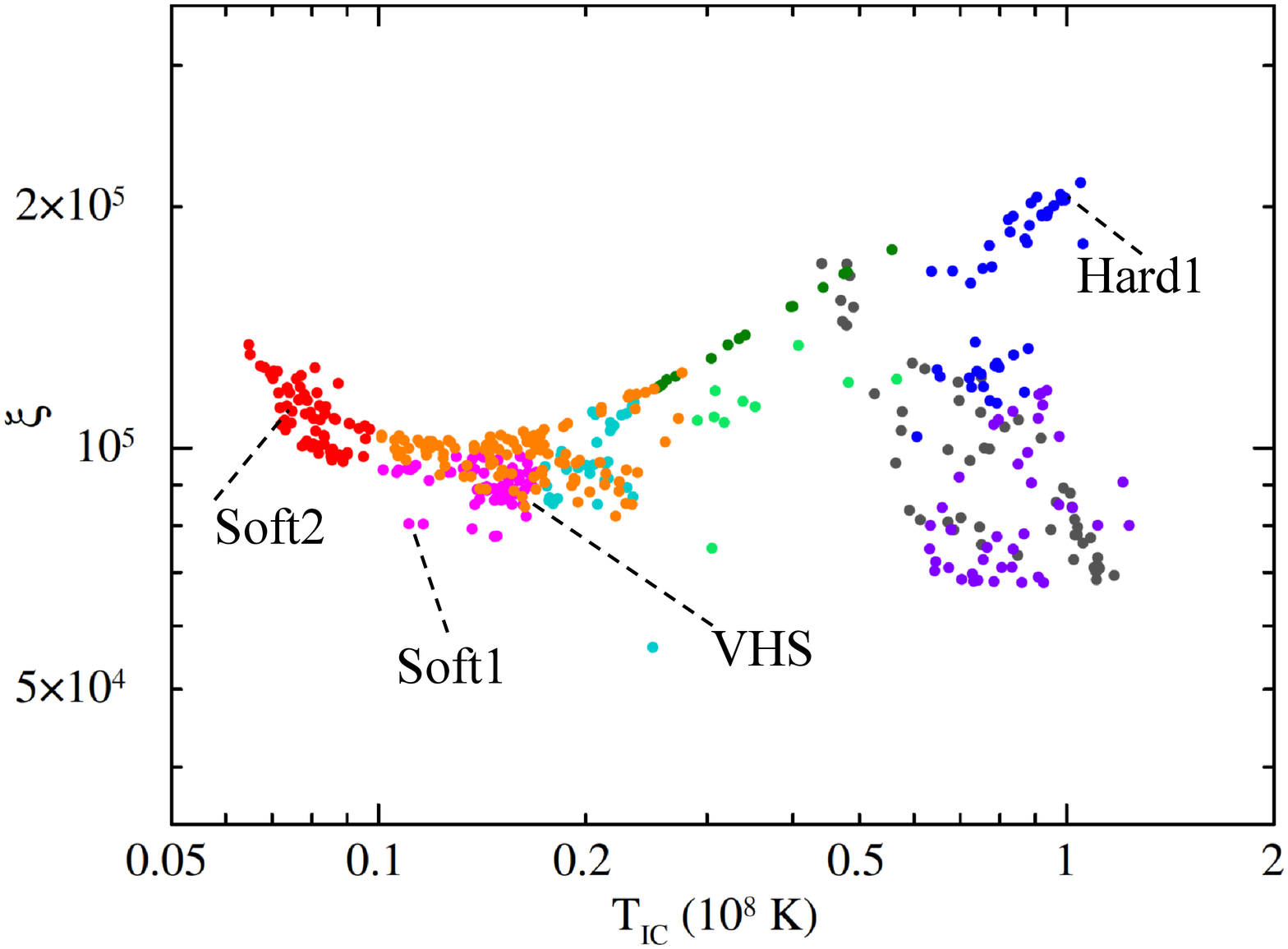}
\caption{(Upper left) Hardness ratio versus luminosity diagram, 
for outbursts of H 1743$-$322 from 2003 to 2011.  
HRs are calculated from the unabsorbed 1--8 keV and 8--100 keV 
fluxes, and the luminosity from the unabsorbed 0.01--100 keV flux, 
all estimated from the best-fit continuum model.
The data points are separated into 9 groups and shown in different colors 
so that these could be tracked in the predictions of wind parameters 
in the other panels. (Upper right) Relation between $T_\mathrm{IC}$ and 
$L/L_\mathrm{Edd}$. (Lower left and right) Evolution of $N_\mathrm{H}$ 
and $\xi$ of thermal winds with respect to $T_\mathrm{IC}$, respectively, 
predicted by the D18 model. The same color codes as the upper 
panels are used. The \chandra\ epochs before 2011 are indicated 
with dashed lines. The 2015 epoch (Hard2) is also indicated in the top left 
panel, by assuming that it is located at the same position as the RXTE point 
with the closest flux and spectral shape in 3--9 keV.
\label{fig:wind_color_code}}
\end{figure*}

The blackbody seen in the low/hard state spectrum in Hard1 
is quite hot and dim. Its temperature ($T_\mathrm{in}$) 
is higher than that in the 2003 epochs, which is inconsistent 
with that expected decrease in disk temperature 
from high/soft state to the low/hard state. 
When we fit the two data simultaneously linking 
\nh~and allowing it to vary, $T_\mathrm{in}$ in Hard1 decreases 
to $\approx 1.2$ keV but still comparable to that of the 2003 epochs. 
We suggest that the thermal component in Hard1 likely 
does not represent the true disk component, but rather is 
compensating for an additional soft Comptonization component
seen in the bright low/hard state \citep[e.g.,][]{mak08,yam13,shi14,mah19}.
We note that in the low/hard state the spectral shape 
below $\sim 1$ keV does not affect the derived wind parameters as 
$T_\mathrm{IC}$ is more sensitive to the hard tail than to a weak disk component.
We also fit the low/hard state spectrum 
with a single power-law model, but the resultant values of 
$L$, \tic, and wind parameters calculated in 
Section~\ref{sec:wind_par} only changed by 10--20\% from
the values in Table~\ref{tab:pars}, which does not affect 
the results of the XSTAR simulation in Section~\ref{sec:simulation}.

\section{Overall Properties of Thermal Wind} \label{sec:wind_par}

Now that we have Compton temperatures in each \rxte~pointed
observation, we can apply the D18 model to calculate the basic 
observable quantities of the thermal wind from the assumed 
system parameters of H 1743$-$322, and hence study the predicted 
evolution of wind properties during the specific outbursts seen here. 

D18 uses the analytic approximation of the 
wind mass-loss rate $\dot{M}_\mathrm{out}$ as a 
function of $L$, derived by \citet{beg83}, with 
two dimensional density structure based on the 
results of the hydrodynamic simulations of \citet{woo96}. 
Assuming a simple density structure, $n(R, i) 
\propto R^{-2} (1-\cos i)$, the column density of 
a thermal wind is derived as 
\begin{equation}
    N_\mathrm{H}(i) = 
    \frac{\dot{M} (1 - \cos i)}{4 \pi R_\mathrm{wind} v_\mathrm{wind} m_\mathrm{I}}, 
    \label{eq:nh}
\end{equation}
and the ionization parameter as 
\begin{equation}
    \xi(i) = \frac{L}{n R_\mathrm{wind}^2} = 
    \frac{4 \pi L v_\mathrm{wind} m_\mathrm{I}}{\dot{M} (1 - \cos i)}, 
    \label{eq:xi}
\end{equation}
where $v_\mathrm{wind}$ is the wind velocity, for which 
the mass-loss averaged sound speed is adopted, 
and $m_\mathrm{I}$ is the mean ion mass for one 
electron ($m_\mathrm{I} \sim 2 \mu$). 

The actual wind launching radius $R_\mathrm{wind}$ is determined 
from that derived from the Compton temperature $R_\mathrm{IC}$.
When the luminosity approaches the Eddington luminosity~\ledd, 
the radiation pressure reduces the effective gravity, leading 
a decrease in the wind launching radius. To consider this effect, 
we adopt a simple correction of $R_\mathrm{IC}$, following D18, as 
\begin{equation}
\overline{R_\mathrm{IC}} = R_\mathrm{IC} (1-\frac{L}{L_\mathrm{Edd}/\sqrt{2}}). \label{eq:lcorr}
\end{equation}
We note that this correction is applicable only 
below $L = L_\mathrm{Edd}/\sqrt{2}$ so 
that the radius is a positive value (see also Section~\ref{sec:discussion} for the limitations 
in the D18 model).
The $R_\mathrm{wind}$ value is given as $R_\mathrm{wind} = 0.2 \overline{R_\mathrm{IC}}$ for 
$L \geq L_\mathrm{crit}$, and $R_\mathrm{wind} = 0.2 \overline{R_\mathrm{IC}} L_\mathrm{crit}/L$ 
for $L < L_\mathrm{crit}$, where the critical luminosity, $L_\mathrm{crit} \sim 3 \times 10^{2} 
T_\mathrm{IC}^{-1/2} L_\mathrm{Edd}$, is defined by the 
luminosity at which the heating rate is sufficient to 
raise the gas temperature to $T_\mathrm{IC}$ at 
$0.2R_\mathrm{IC}$ so that it can escape \citep{beg83}. 
Thus, the basic wind parameters \nh, $\xi$, $R_{\rm wind}$ and $v_\mathrm{wind}$ can be estimated from $L$ and $T_\mathrm{IC}$
given the assumed system parameters $M_\mathrm{BH}$, $i$, 
and $R_\mathrm{out}$. 
Table~\ref{tab:pars} lists the wind parameters estimated 
from the D18 model at the \chandra\ epochs for this system.

To understand how the properties of thermal winds 
change in an outburst, we plotted the three observable 
parameters, $L_\mathrm{X}$, $N_\mathrm{H}$, and $\xi$ 
with respect to \tic\ in Figure~\ref{fig:wind_color_code}. 
We also included the hardness versus luminosity diagram, 
so that we can easily associate these parameters with 
spectral states. 
The launch radius of the wind is generally $0.2R_\mathrm{IC}\propto 
1/T_\mathrm{IC}$, while $v_\mathrm{wind}\propto T_\mathrm{IC}^{1/2}$. 
Hence, $N_\mathrm{H} \propto L T_\mathrm{IC}^{1/2}$ while $\xi \propto T_\mathrm{IC}^{1/2}$. 
At highest luminosities above 30--40\%~\ledd, however, $N_\mathrm{H}$ and $\xi$ 
become even larger and lower, respectively, due to the effect of the radiation pressure correction.

Since \tic\ increases as the X-ray spectrum becomes 
harder, the \tic\ versus $L_\mathrm{X}$ plot 
(the top right panel in Fig.~\ref{fig:wind_color_code}) 
can be regarded as the hardness-intensity diagram. Indeed, 
it makes almost the same track as the hardness 
luminosity diagram (top left in Fig.\ref{fig:wind_color_code}) 
and shows hysteresis; the transition from the low/hard state to 
the high/soft state occurs at a higher luminosity 
than the opposite transition. A similar track 
can be seen in the \tic-\nh\ plot, as $N_\mathrm{H} \propto L_\mathrm{X} T_\mathrm{IC}^{1/2}$ 
so it depends more strongly on $L_\mathrm{X}$ than spectral hardness. 
By contrast, the \tic-$\xi$ plot exhibits a very different track, as $\xi\propto T_\mathrm{IC}^{1/2}$ only. 

We note that the \tic-$\xi$ diagram does not directly indicate 
the visibility of the Fe K absorption lines, because the $\xi$ value is 
estimated from the bolometric luminosity, and does not incorporate the information on the 
spectral shape. 
The hard X-ray fraction in the total luminosity is $\sim 1$ order 
of magnitude larger in the low/hard state (coloured in black, purple, and blue 
in Fig.~\ref{fig:wind_color_code})
than in the high/soft state (coloured in pink, red, and orange).
Hence, the wind is completely ionized during the former state, leading 
to the absence of the lines, whereas it often produces lines in the latter state, 
even if the $\xi$ values are not very different (see also Section~\ref{sec:discussion}). 
We incorporate this spectral shape information in the next section.

\section{Detailed Photoionized Plasma Simulations} \label{sec:simulation}

Adopting the wind parameters given in Table~\ref{tab:pars} 
as input to the XSTAR photoionization code, we made
detailed simulations of the wind absorption features 
at the \chandra~epochs. 
We used XSTAR version 2.41 together with XSTAR2XSPEC, which 
runs XSTAR simulations multiple times to provide an XSPEC 
table model of ionized absorption, based on the simulation 
results. The XSTAR simulations were performed 
for the individual \chandra\ epochs, using their 
best-fit continuum models as the input SEDs. 
Here, the density $n_0$ at the wind launching radius 
and the ionizing luminosity $L_0$ in 0.0136--13.6 keV 
used in XSTAR were fixed at the values in 
Tab.~\ref{tab:pars}, while $\xi$, \nh, and the blueshift 
(or the line-of-sight velocity $v_\mathrm{wind}$)
were varied. The turbulent velocity was set at 
300 km s$^{-1}$ and abundances were set to solar. 
In these simulations, we assumed that 
the density of the ionized plasma is constant 
with respect to radius, 
although the D18 model adopts the radial dependence as 
$n(r) = n_0 (r/R_\mathrm{wind})^{-2}$. This is because 
the simulations never converge when we use the latter 
dependence, due to technical reasons in 
XSTAR\footnote{\url{https://heasarc.gsfc.nasa.gov/xstar/docs/html/xstarmanual.html}}. 

The resultant table model for the individual epochs 
was added to their best-fit continuum models obtained 
from the fits to the broadband data (see Sec.~\ref{sec:xrtana})
and applied to the \chandra/HETGS data at each epoch.
We utilized first order HEG 
spectra in 3--9 keV and their response files, 
downloaded from the Chandra Transmission Grating 
Data Archive and Catalog \citep{hue11}. 
The HEG continuum spectra 
were found to be somewhat harder than the corresponding 
\rxte/PCA spectra, and significant residuals remain
mainly above $\sim 8$ keV. 
This could be be due to time variability 
between the \rxte\ and \chandra\ observations, 
which are not exactly simultaneous, or spectral 
distortion by dust scattering halo \citep{all18}, or a 
calibration uncertainty in \chandra\ responses.
To reduce the discrepancy between the HEG data 
and the continuum model, we varied $T_\mathrm{in}$ 
and normalization of {\tt diskbb} for the high/soft state 
and $\Gamma$ and $F_\mathrm{scat}$ of the {\tt simpl} model 
for the low/hard state and the very high state, 
We note that this treatment, which allows the 
above parameters to be different from those 
obtained from the \rxte(+\swift/BAT) data, 
only slightly changes \tic\ from the original values 
and does not affect the wind parameters.

In the following, we show the results of the simulations 
and demonstrate how they reproduce the data at each epoch. 

\subsection{High/soft State (Soft1 and Soft2)}

\begin{figure}[htb]
\epsscale{1.1}
\plotone{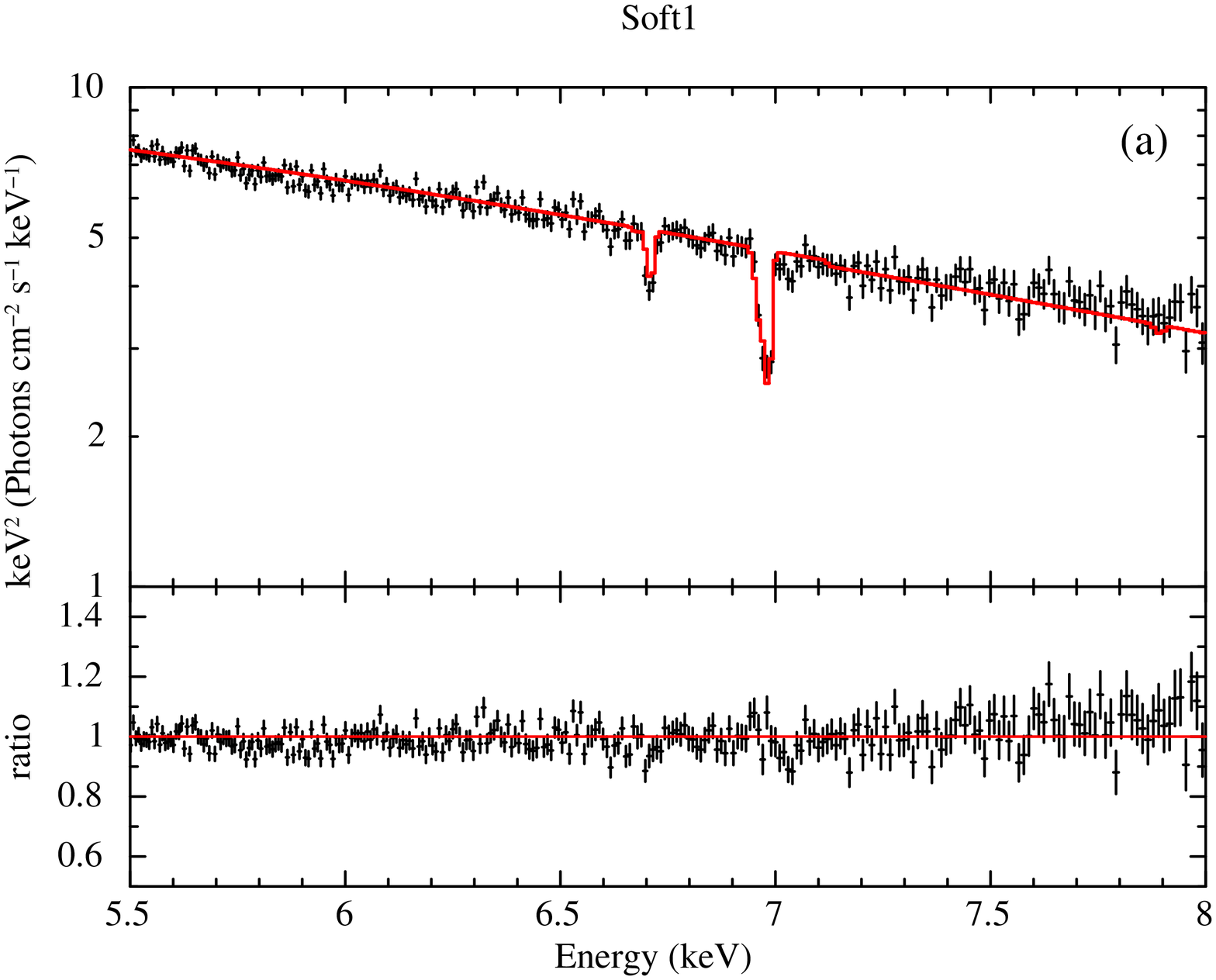}
\plotone{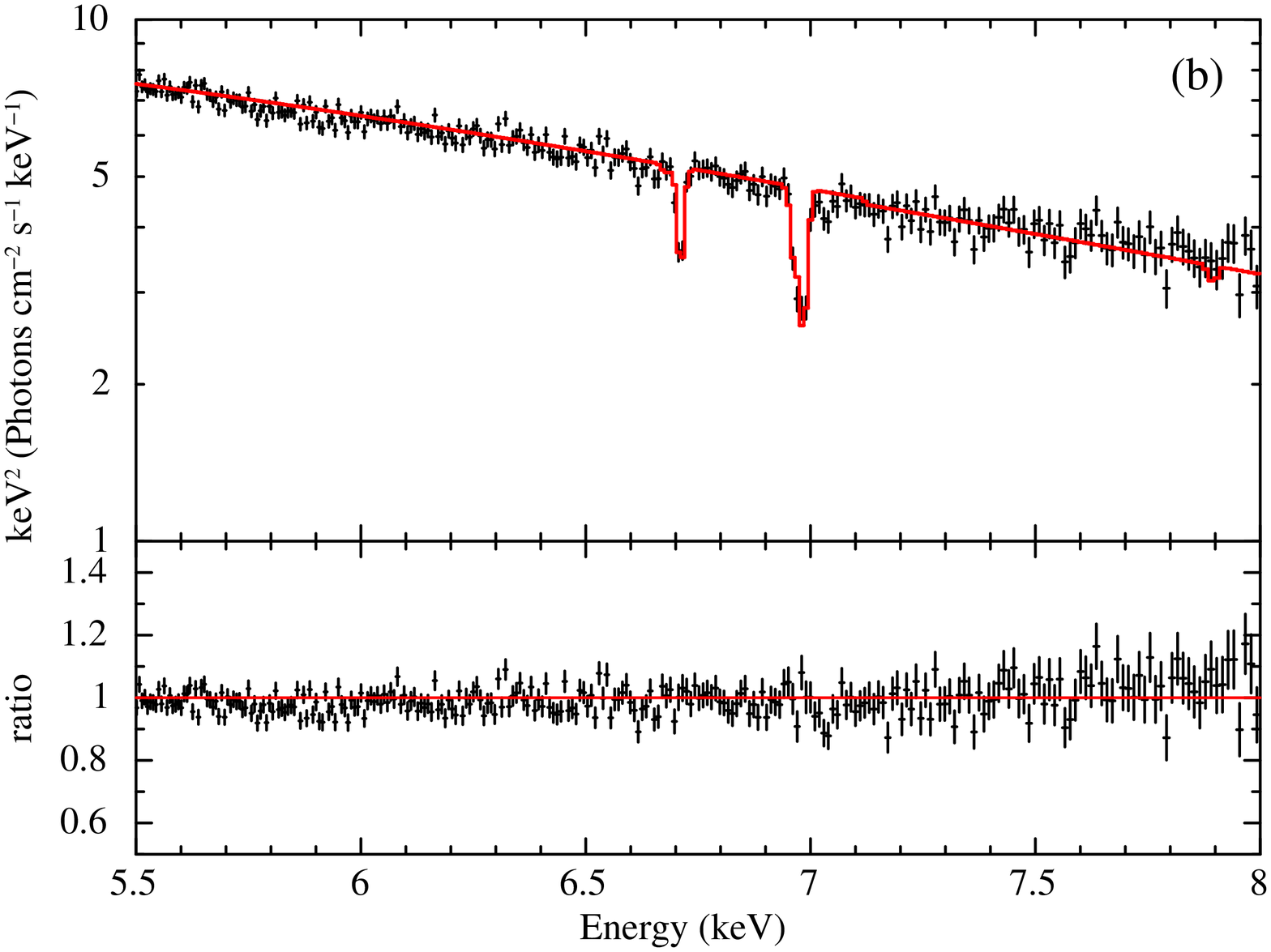}
\caption{Comparison between the Chandra HETGS unfolded data 
at Soft1 and simulated absorption line spectrum based on the D18 model. 
The best-fit model obtained in Section~\ref{sec:xrtana} 
was adopted as the continuum model. 
(a) $N_\mathrm{H}$ and $\xi$ are fixed at the value predicted 
by the wind model (see Table~\ref{tab:pars}). 
(b) The best-fit model obtained by by allowing $N_\mathrm{H}$, 
$\xi$, and $v_\mathrm{out}$ to vary (see text).
The lower panels present the data versus model ratios.  
\label{fig:sim_hss}}
\end{figure}

\begin{figure}[htb]
\epsscale{1.1}
\plotone{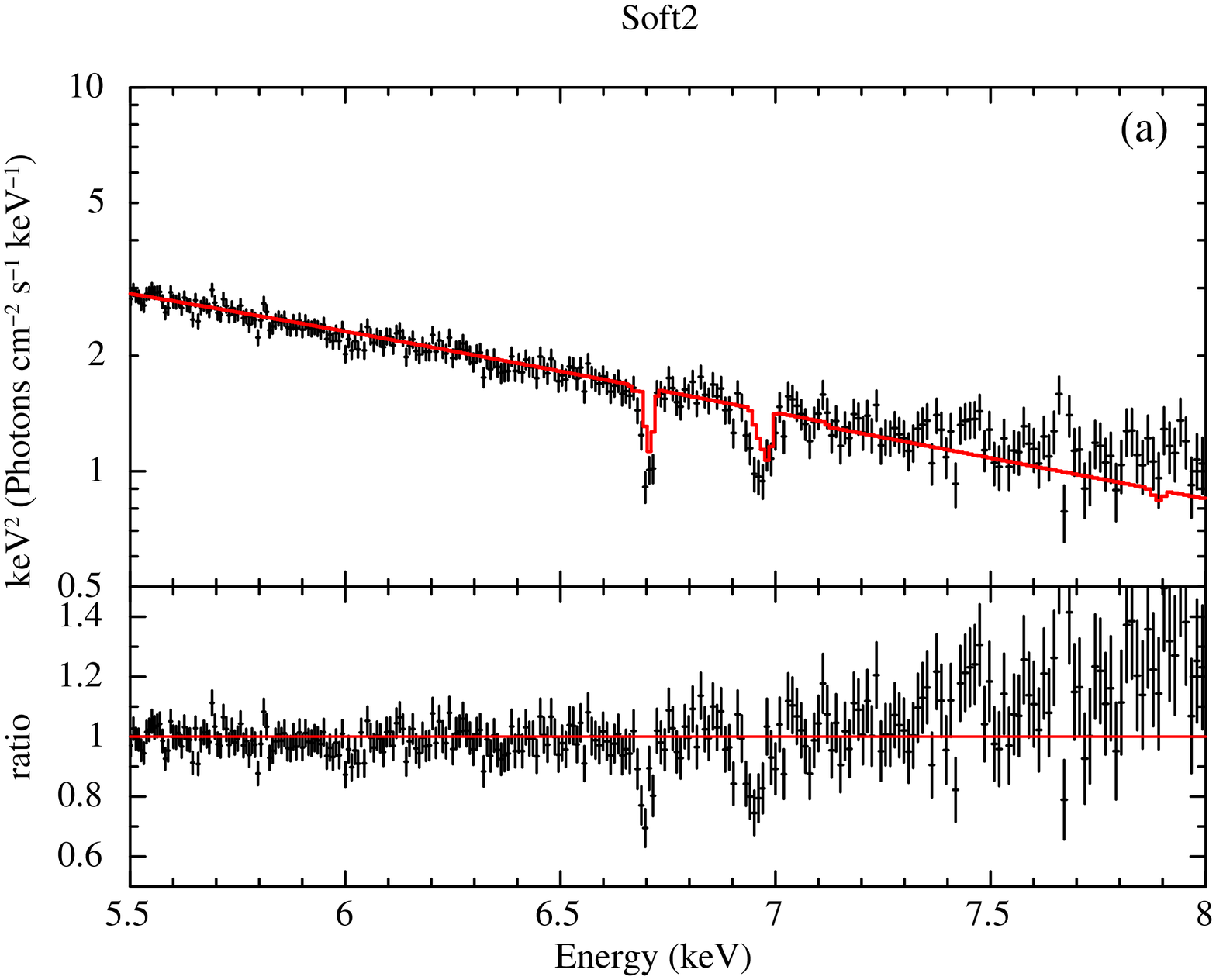}
\plotone{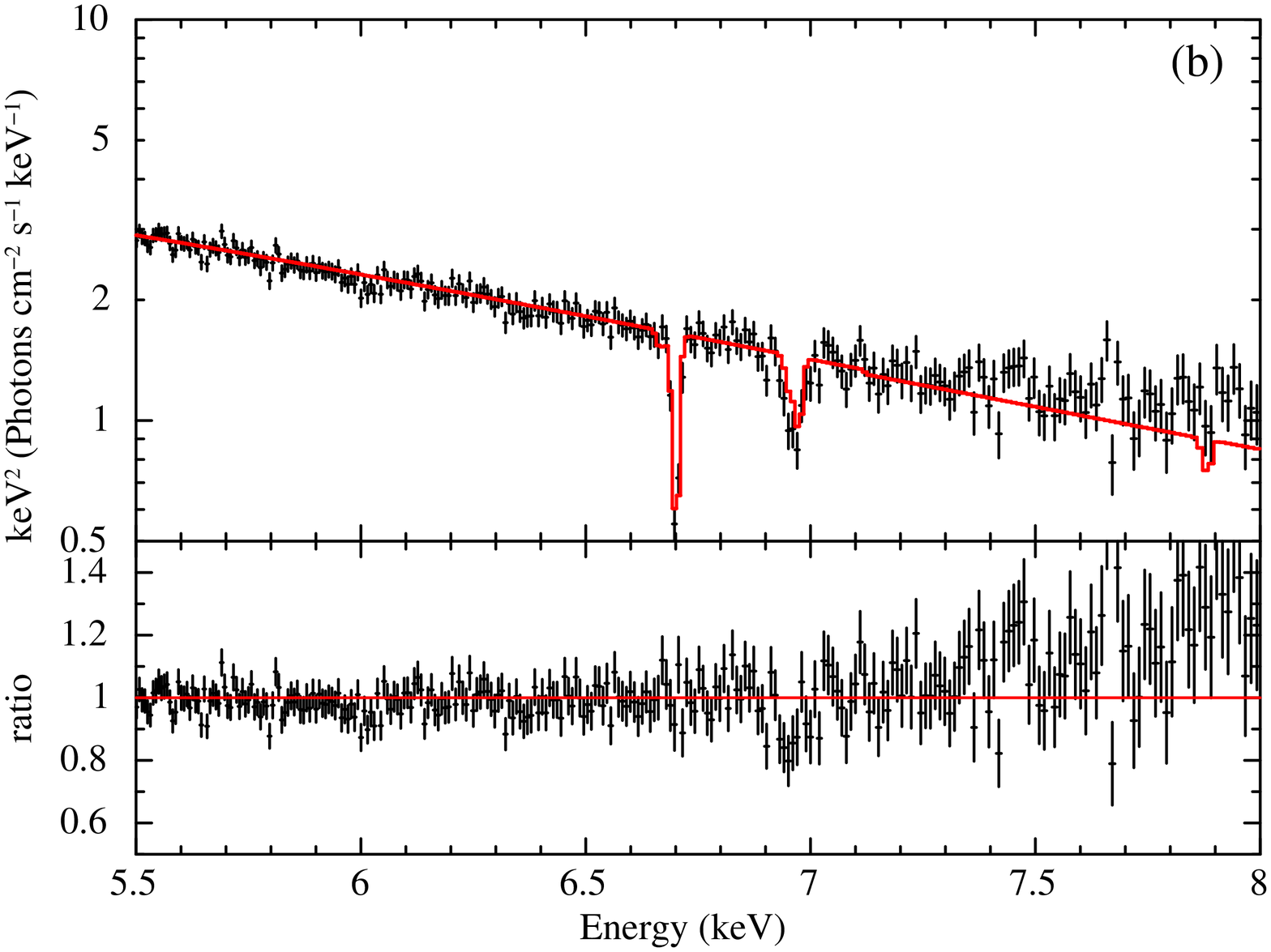}
\caption{Same as Fig.~\ref{fig:sim_hss} for the 
high/soft state data at Soft2.
\label{fig:sim_uss}}
\end{figure}

In Figure~\ref{fig:sim_hss}(a), the model obtained from the 
XSTAR simulation is compared with the \chandra\ spectrum 
for Soft1. We first fix \nh, $\xi$, and the 
blueshift velocity at the values in Table~\ref{tab:pars}.
The observed He-like and H-like Fe lines at 6.7 keV and 
7.0 keV, respectively, are well reproduced by the model. 
By contrast, Fig.~\ref{fig:sim_hss}(b), allows the 
three wind parameters to vary, to find the best fit 
description of the data. In this case, the fit quality 
marginally improved from the case of fixed wind parameters, 
from $\chi^2/$dof $= 1774/1096$ to $1761/1093$, 
and $N_\mathrm{H} = (4 \pm 1)
\times 10^{22}$ cm$^{-2}$, 
$\xi = 3^{+2}_{-1} \times 10^4$ erg cm s$^{-1}$, and 
$v_\mathrm{wind} = 5 \pm 1 \times 10^2$ km s$^{-1}$ 
were obtained. This combination gives very similar line
equivalent widths as the material is so highly ionized that the
decrease in ionization parameter means that less of the iron is completely ionized, so
increases the column in FeXXV and XXVI in such a way as to offset 
the decrease in overall column density. Whichever combination is chosen, 
it is clear that the thermal wind model predictions can explain 
this observation within a factor of $\sim$2 uncertainties. 

The D18 analysis had the observed source $L/L_{Edd}=0.1$ at Soft1
due to the difference in distance/mass/spin, but assumed limb darkening 
so that their intrinsic $L/L_{Edd}=0.3$--$0.4$ as assumed here from the 
observed spectrum at these different system parameters. 
Their estimate for $N_\mathrm{H}=8\times 10^{22}$~cm$^{-2}$ for a 
source at this luminosity is slightly larger than the 
$N_\mathrm{H}=6.7\times 10^{22}$~cm$^{-2}$ predicted here due to their larger $R_{out}$.

Figure~\ref{fig:sim_uss}(a) and (b) compare the Soft2 data taken 
at $L/L_{Edd}=0.2$ and the corresponding XSTAR absorption model, 
in the same way as Soft1. At this epoch the source exhibited a much 
softer SED and had a hard tail $\sim 10$ times weaker than Soft1.
Our prediction using the D18 model somewhat underestimates 
the wind column density and thereby the Fe line strengths (Fig.~\ref{fig:sim_uss}a).
When the wind parameters were allowed to vary, the 
chi-squared value was significantly reduced from 
$\chi^2/$dof $= 1406/1234$ to $1343/1231$ and the discrepancy between 
the data and model was mitigated (Fig.~\ref{fig:sim_uss}b). The best-fit 
absorption model gives $N_H= 2.6^{+0.8}_{-2.1} \times 10^{22}$~cm$^{-2}$, 
$\xi = 8.8^{+0.5}_{-0.3} \times 10^4$ erg cm s$^{-1}$, 
and $v_\mathrm{wind} < 2 \times 10^2 $ km s$^{-1}$. 
Thus, the thermal wind model can again explain, within a factor of $\sim$2, 
the absorption features in the high/soft state spectrum with a very weak 
hard tail.

\subsection{Low/hard State (Hard1 and Hard2)}

\begin{figure}[htb]
\epsscale{1.1}
\plotone{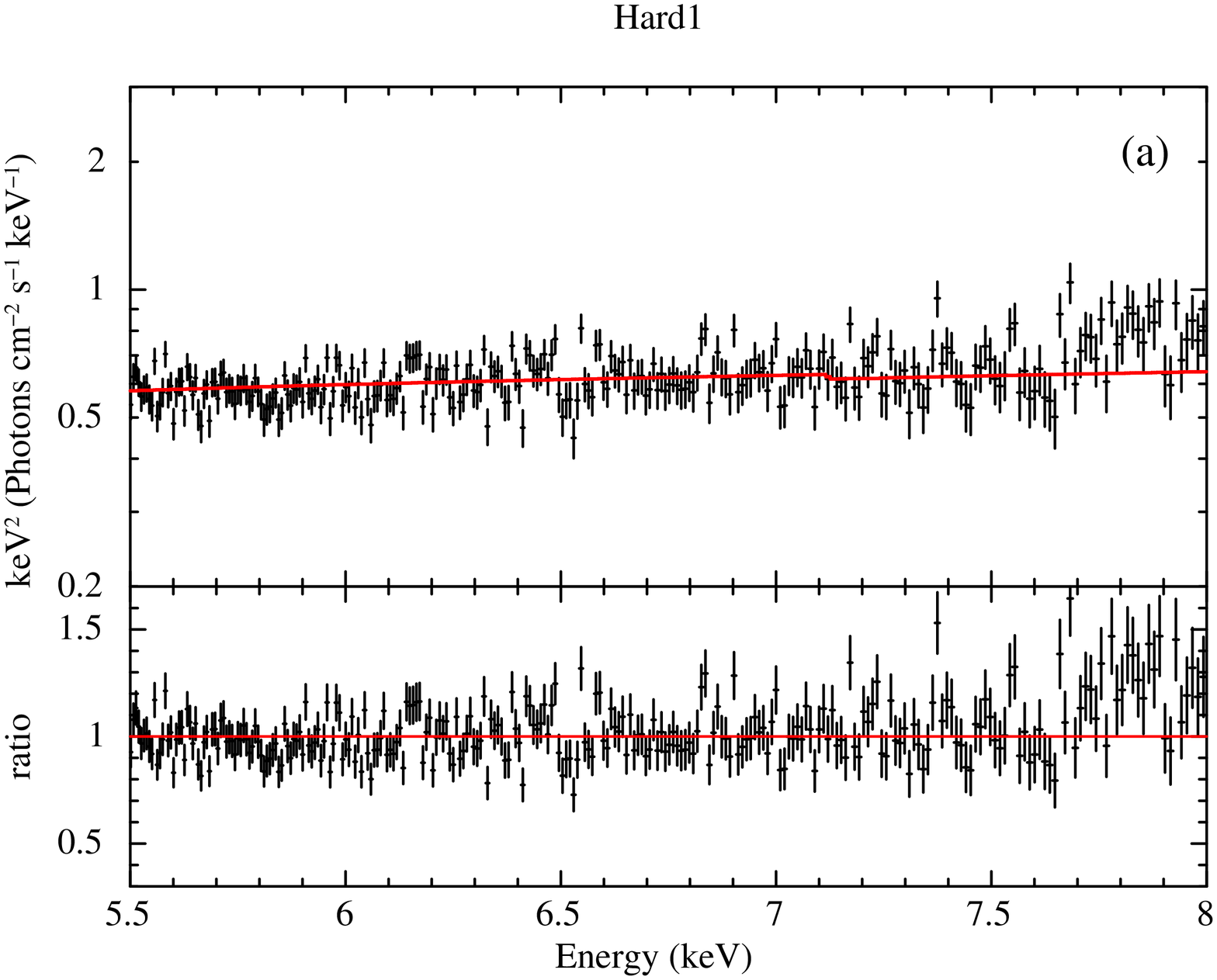}
\plotone{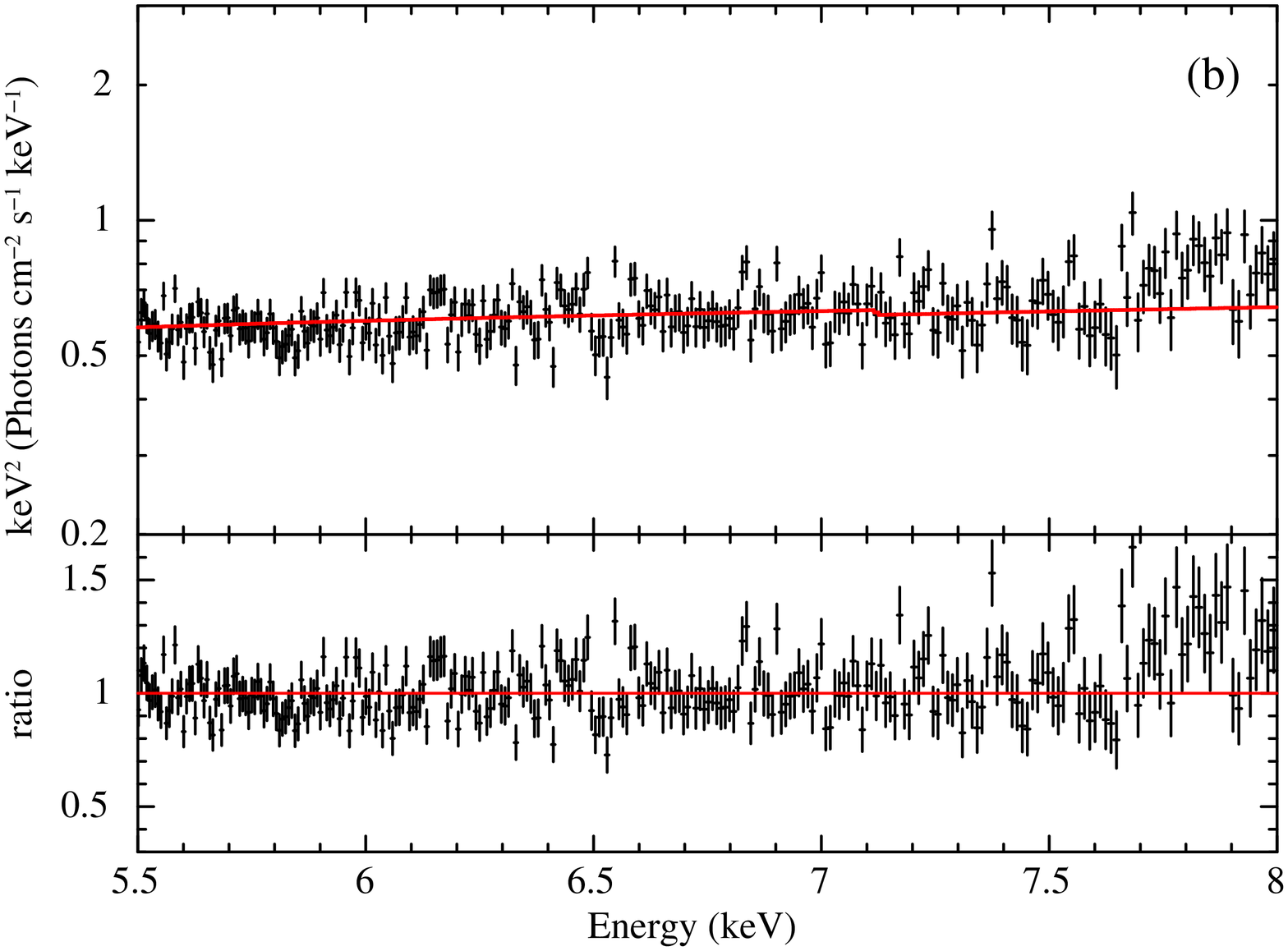}
\caption{Same as Fig.~\ref{fig:sim_hss} for the 
low/hard state data at Hard1, where (a) $N_\mathrm{H}$ and 
$\xi$ are fixed at the value predicted by the wind model  
and (b) a factor of 2 smaller values are adopted.
\label{fig:sim_lhs}}
\end{figure}

\begin{figure}[htb]
\epsscale{1.1}
\plotone{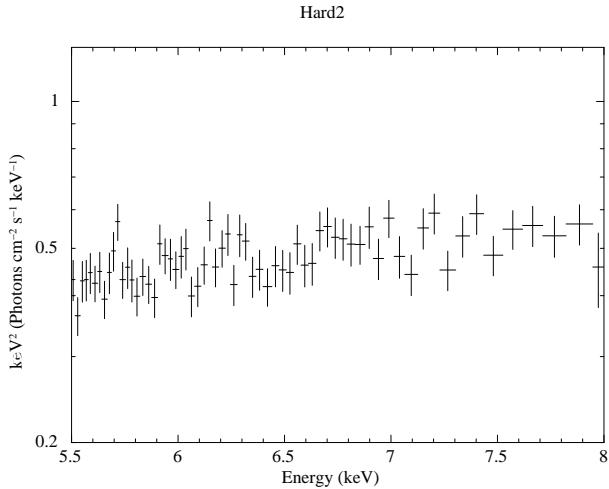}
\caption{\chandra\ HETGS spectrum around 7 keV at Hard2. Data from three observations are co-added to improve statistics (see Tab.~\ref{tab:obslist_chandra} for more details).
A power-law model with a photon index of 1.3 was used to plot the unfolded spectrum.
\label{fig:sim_lhs2}}
\end{figure}

Figure~\ref{fig:sim_lhs} shows the same sequence of fits to 
the \chandra~data at Hard1. The upper panel shows the predicted
absorption spectrum for the wind parameters fixed at the
predicted values in Table~\ref{tab:pars}. There are no significant 
features, which matches well to the observed data. The lower panel 
shows the resulting wind scaled in the same way as the best fit 
to Soft1 in the high/soft state; 
i.e. we reduce the column density and 
ionization parameter by a factor of 2 from the model predictions, 
but the wind is still not visible. We note that the result unchanged 
even when we increase the column by a factor of 2 following the fit to 
Soft2.

Thus thermal wind model tailored to the observed luminosity and SED predicts 
no significant Fe K absorption lines in the low/hard state, consistent with 
the \chandra\ observation. This does not mean that the
wind has disappeared.
The simple thermal wind models predict that this lower luminosity spectrum 
should have a column which is only a factor 3 smaller than that seen in 
the high/soft state. However, the higher $T_\mathrm{IC}$ means that the wind 
is now launched from much closer in. The ionization state is higher 
so the column of FeXXV and even FeXXVI is too small to be observed. 
The wind not only responds via photoionisation to the changing spectral shape 
(see e.g \citealt{cha13}), but also responds in terms of its 
launch radius, velocity and density due to the change in Compton temperature.

This is similar to the conclusion of D18, though they 
had an inferred $L/L_\mathrm{Edd}=0.02$ due to the difference in distance/mass/spin, 
rather than the $L/L_\mathrm{Edd}=0.06$ determined here at these different system 
parameters. This is a bright/low hard state seen on the fast rise, where 
the transition to the high/soft state can occur at much higher $L/L_\mathrm{Edd}$ 
than the typical transition value of $0.02 L_\mathrm{Edd}$ seen on the slow decline 
(hysteresis). Our estimate of $N_\mathrm{H} \sim 2.6 \times 10^{22}$ cm$^{-2}$ is 
higher than the $N_\mathrm{H} = 1.8 \times 10^{22}$~cm$^{-2}$ of D18 for these data, 
as the higher source luminosity is more than offsetting the effect of a smaller 
outer disk radius.

In Figure~\ref{fig:sim_lhs2} we also show the HETGS spectrum around 7 keV obtained at Hard2 in 2015, where the source was a factor of $\sim 2$ fainter than Hard1 (i.e., $L/L_\mathrm{Edd} \sim 0.03$). 
No significant lines are visible, like Hard1.
Although this epoch is out of the coverage of our calculation with D18 model, the wind parameters should be almost the same as those of Hard1, considering the only factor-of-two difference in flux. Our prediction is hence no lines in this fainter low/hard state, which is again consistent with the observation.

\subsection{Very High State (VHS)}

\begin{figure}[htb]
\epsscale{1.1}
\plotone{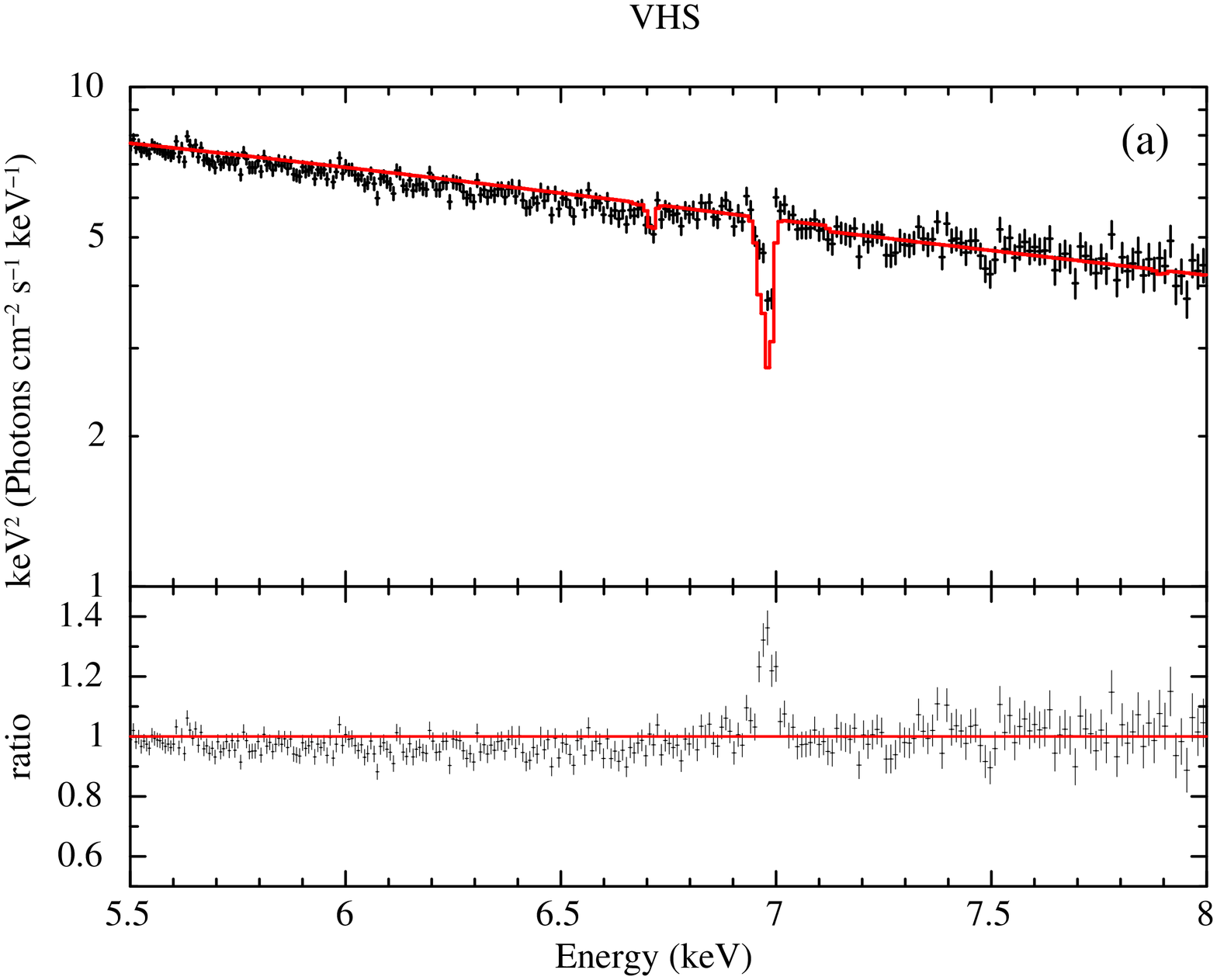}
\plotone{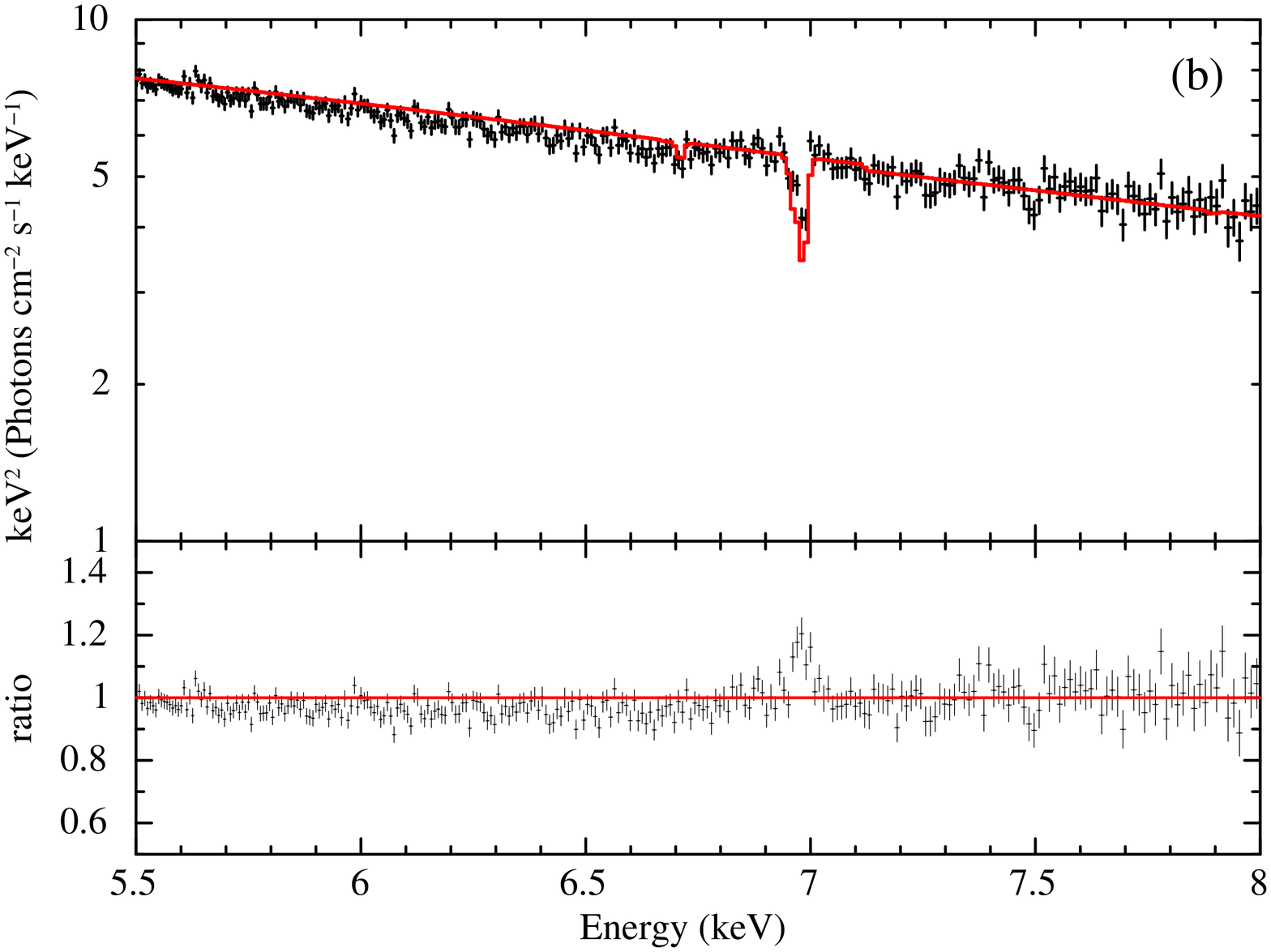}
\plotone{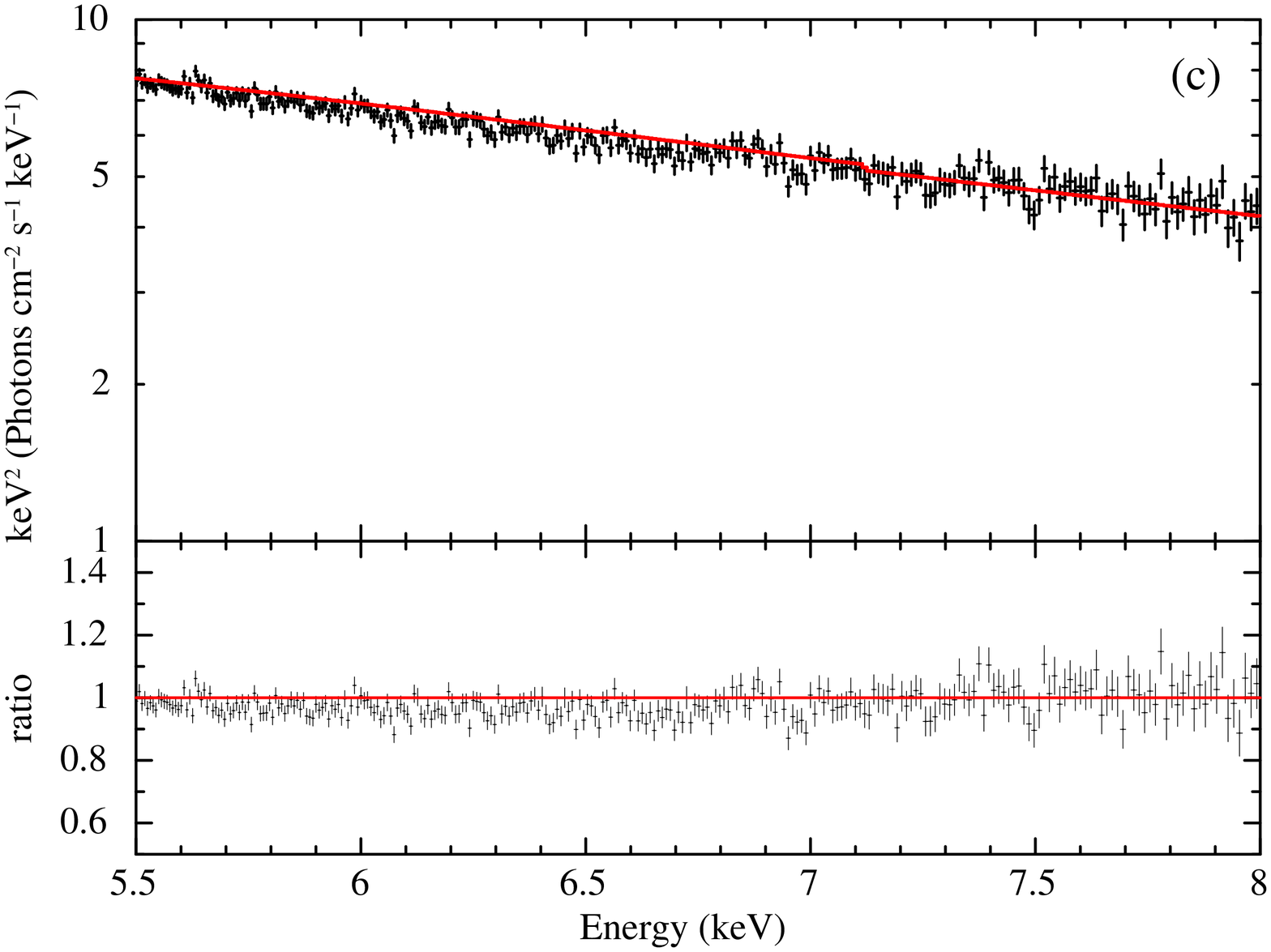}
\caption{Same as Fig.~\ref{fig:sim_hss} for the 
data at VHS, where (a) $N_\mathrm{H}$ and 
$\xi$ are fixed at the value predicted by the wind model, 
(b) a factor of 2 smaller values are adopted. The unfolded 
spectra made with XSPEC are model-dependent and panel (a) and 
(b) are actually affected by the adopted absorption line models.
To avoid this effect, panel (c) presents the unfolded spectrum 
to which only continuum model is applied. \label{fig:sim_vhs}}
\end{figure}

Figure~\ref{fig:sim_vhs} makes a comparison of the 
VHS data in the very high state and the corresponding 
XSTAR simulation result, in the same manner as the other 
epochs. The HETGS spectrum shows no significant lines, 
although there may be a hint of a weak Fe XXVI line at 7 keV 
(see Fig.~\ref{fig:sim_vhs}c). 
Using D18 model we obtained a large wind column, 
$1.2 \times 10^{23}$ cm$^{-2}$, and a moderate 
ionisation parameter, $8.5 \times 10^4$ erg cm s$^{-1}$, 
and thus our XSTAR simulation predicts significant 
detection of the Fe XXVI line, which is inconsistent 
with the observation. This discrepancy is not changed
even if we consider the factor-of-two uncertainty in 
the wind parameters found in the high/soft state.

Instead, we consider the more detailed thermal wind structure
derived by \citet{beg83b} (see also \citealt{ost91,tom19}). These 
papers analyze the vertical structure of the X-ray irradiated upper layer 
of the disk. In the original paper of \citet{beg83}, the 
Compton heated material forms a static atmosphere over the inner disk. 
It is very easy for this to go optically thick in directions along the 
equatorial plane, shielding the outer disk from illumination until 
the convex disk shape brings the disk surface out of the shadow. 
\citet{tom19} show that this first directly illuminated 
point on the outer disk is almost exactly at $R_\mathrm{out}$ 
for the high/soft state of H~1743$-$322 (Soft1). We use 
their equations for the very high state parameters here and find
that the higher Compton temperature means that the inner atmosphere 
has a larger scale height, so casts a longer shadow, shielding the 
disk from direct irradiation across its entire extent (out to $1.4R_\mathrm{out}$). 
The precise suppression of illuminating flux depends on the detailed 
vertical structure of the inner disk atmosphere and X-ray corona 
geometry in this state, but the wind properties give a potential 
observable diagnostic of these poorly known quantities.

\section{Discussion and Conclusions} \label{sec:discussion}

Using the D18 model and X-ray data of H 1743$-$322, 
we have investigated how thermal winds should evolve 
over an entire period of an outburst. The high cadence 
and the wide energy coverage of \rxte\ and \swift/BAT 
enabled us to accurately estimate the Compton 
temperature \tic\ across all the outbursts. These broadband 
spectra then enabled us to predict the observable parameters 
of thermal winds throughout the outburst cycles.

We also make detailed photoionization models of the 
predicted thermal winds to compare with \chandra\ high 
resolution spectral epochs. These match very well to 
the observed properties of the wind in the high/soft state data
at $\sim 30$\% $L_\mathrm{Edd}$ (Soft1) 
and at $\sim 20$\% $L_\mathrm{Edd}$ (Soft2). 
The \nh\ and $\xi$ values derived directly from 
the D18 model differ only by a factor of 
$\sim$2 from the best-fit result. 
This strongly suggests that the thermal 
driving is the main launching mechanism 
of the observed wind, at least in this state. 
There is very little room in the data for any substantial 
contribution from a magnetic wind. 

The corresponding prediction for the bright low/hard 
state at 3$\sim$6\% $L_\mathrm{Edd}$ is that the 
absorption lines should not be visible, and this is again 
consistent with the \chandra\ data. 
The models predict that the harder spectrum can launch a wind
from closer in, so even though the predicted column decrease 
is only a factor of 3, its typical ionization parameter is increased 
by a large factor, especially when considering only the ionization 
of the iron species. These are controlled by X-rays above 
$\sim$8 keV where the difference in $\xi$ becomes much 
larger than that estimated from the bolometric luminosity. 
In the high/soft state, the X-ray flux is dominated by the 
direct disk component below $\sim$10 keV, and the contribution 
of the hard tail is only $\sim 7$\% in the total 
luminosity at Soft1 and $\sim 0.2$\% at Soft2, 
whereas in the low/hard state, the hard X-rays  
above 8 keV contributes $\sim 75$\%. 
Thus, the ionization parameter for the
ionization of iron is about 30--1000 times higher, 
and iron is almost completely ionized in 
the low/hard state.

Even though our predicted column is not visible even 
with \chandra\ in the low/hard state, it is still an 
overestimate of the spectral features, as the inner disk 
heated atmosphere can shield the outer disk from 
illumination \citep{beg83b,tom19}. The 
larger scale height of this inner disk atmosphere 
leads to an increased shadow across the outer disk, 
predicting even lower wind mass loss rates in 
the low/hard state \citep{tom19}.

We may be seeing evidence of this shadow at highest luminosities,
during the very high state. Wind models without the shadow predict that the
highest column density should be seen in these intermediate hardness spectra
(the turquoise points in Fig.~\ref{fig:wind_color_code}). The VHS 
Chandra data are close to this branch, and the photo-ionization 
simulations of the column and ionization state predicted by the
simple thermal models of D18 give features which should be  
easily observable in the data, yet are not detected. 
We note, however, that the estimated wind 
parameters above 30--40\%~\ledd, where the 
radiation pressure effect plays an important role, may 
include an additional large systematic error, because 
our radiation pressure correction is only a simple 
approximation; we only considered the decrease of 
the wind launching radius, but the density and 
velocity structures would also change as well 
(D18, \citealt{tom19}), which is ignored in our 
assumption.

Our calculation using the D18 model is based 
on simple assumptions, and contains uncertainties 
caused by the system parameters including 
the disk size, the black hole mass, inclination, 
and the distance, the geometry (and hence illumination as a function of angle) 
of the X-ray source, and shape of the streamlines 
in the thermal wind, especially at high 
luminosities.
All of them can affect the results, even though the Compton 
temperature was directly estimated from 
the actual X-ray spectra. More precise models
require better determination of the system parameters, 
coupled to full radiation hydrodynamics to calculate 
the 2-dimensional structure of the wind 
streamlines, followed by detailed radiation 
transfer to produce the spectral features 
\citep{tom19}. Nonetheless, even our simplified 
thermal wind model can already explain the observed 
behavior of the absorption lines in the low/hard 
and high/soft states. There is very little room 
for a strong magnetic wind which is not completely 
ionized in these data.
Thermal winds do however over-predict the lines 
in the very high state. While this could be some form 
of magnetic suppression of the wind \citep{wat18}, it 
seems more likely that this is due to an increasing 
scale height of the inner disk atmosphere 
reducing X-ray irradiation of the outer disk, where the thermal winds are launched.
The wind features (or lack of them) could then give 
insight into the poorly constrained vertical structure 
of the X-ray source and X-ray illuminated inner accretion disk.

\acknowledgments
We thank the anonymous referee for providing valuable comments.  
MS acknowledges support by the Special Postdoctoral Researchers 
Program at RIKEN. This work is partly supported by a Grant-in-Aid 
for Young Scientists (B) 16K17672 (MS). 
This research has made use of MAXI data provided by RIKEN, 
JAXA and the MAXI team and {\it Swift} data supplied by the UK Swift 
Science Data Centre at the University of Leicester. 
CD acknowledges the Science and Technology Facilities Council (STFC) 
through grant ST/P000541/1 for support.

%

\vspace{5mm} 
\facilities{\rxte\ (PCA and HEXTE), \swift\ (BAT), \chandra\ (HETGS)}
\software{XSPEC (v12.9.0n; \citealt{arn96}), HEAsoft (v6.19; HEASARC 2014), XSTAR 
(v2.41; Kallman 1999 on Astrophysics Source Code Library, \citealt{kal01})}


\begin{thebibliography}{}
\bibitem[Allen et al.(2018)]{all18} Allen, J.~L., Schulz, N.~S., Homan, J., et al.\ 2018, \apj, 861, 26
\bibitem[Arnaud(1996)]{arn96} Arnaud, K.~A.\ 1996, Astronomical Society of the Pacific Conference Series, 101, 17
\bibitem[Begelman et al.(1983)]{beg83} Begelman, M.~C., McKee, C.~F., \& Shields
, G.~A. \ 1983, \apj, 271, 70
\bibitem[Begelman \& McKee(1983)]{beg83b} Begelman, M.~C., McKee, C.~F.\ 1983, \apj, 271, 89
\bibitem[Bradt et al.(1993)]{bra93} Bradt, H.~V., Rothschild, R.~E., \& Swank, J.~H.\ 1993, \aaps, 97, 355
\bibitem[Capitanio et al.(2009)]{cap09} Capitanio, F., Belloni, T., Del Santo, M., et al.\ 2009, \apj, 398, 1194 
\bibitem[Coriat et al.(2012)]{cor12} Coriat, M., Fender, R. P., Dubus, G.\ 2012, \mnras, 414, 1991
\bibitem[Chakravorty, Lee \& Neilsen 2013]{cha13} Chakravorty, S., Lee, J., \& Neilsen, J., 2013, \mnras, 436, 560
\bibitem[D{\'{\i}}az Trigo \& Boirin(2013)]{dia13} D{\'{\i}}az Trigo, M., \&  Boirin, L.\ 2013, Acta Polytechnica, 53, 659
\bibitem[D{\'{\i}}az Trigo et al.(2014)]{dia14} D{\'{\i}}az Trigo, M., Migliari, S., Miller-Jones, J.~C.~A., \& 
Guainazzi, M.\ 2014, \aap, 571, 76
\bibitem[D{\'{\i}}az Trigo \& Boirin(2016)]{dia16} D{\'{\i}}az Trigo, M., \&  Boirin, L.\ 2016, Astronomische Nachrichten, 337, 368
\bibitem[Done(2010)]{don10} Done, C.\ 2010, arXiv:1008.2287 
\bibitem[Done et al.(2018)]{don18} Done, C., Tomaru R., \& Takahashi, T.\ 2017, \mnras, 473, 838 (D18)
\bibitem[Fukumura et al.(2014)]{fuk14} Fukumura K., Tombesi F., Kazanas D., et al.\ 2014, \apj, 780, 120,
\bibitem[Fukumura et al.(2017)]{fuk17} Fukumura K., Kazanas D., Shrader C., et al.\ 2017, Nature Astron., 1, 0062
\bibitem[Higginbottom et al.(2014)]{hig14} Higginbottom N., Proga D., Knigge C., et al.\ 2014, \apj, 789, 19
\bibitem[Hori et al.(2018)]{hor18} Hori, T., Ueda, Y., Done, C., Shidatsu, M., \& Kubota,
A.\ \apj, 869, 183
\bibitem[Huenemoerder et al.(2011)]{hue11} Huenemoerder, D.~P., Mitschang, A., Dewey, D., et al.\ 2011, \aj, 141, 129
\bibitem[Ingram et al.(2009)]{ing09} Ingram, A., Done, C., \& Fragile, P.~C.\ 2009, \mnras, 397, 1
\bibitem[Kallman \& Bautista(2001)]{kal01} Kallman, T.\& Bautista, M.\ 2001, \apjs, 133, 221   
\bibitem[Kaluzienski \& Holt(1977)]{kal77} Kaluzienski, L. J., \& Holt
, S. S.\ 1977, IAU Circ.\ 3099
\bibitem[Kotani et al.(2000)]{kot00} Kotani, T., Ebisawa, K., Dotani T., et al.\ 2000, \apj, 539, 413
\bibitem[Kubota et al.(2007)]{kub07} Kubota, A., Dotani, T., Cottam, J., et al.\ 2007 \pasj, 59, 185
\bibitem[Makishima et al.(2008)]{mak08} Makishima, K., Takahashi, H., Yamada, S., et al.  2008, \pasj, 60, 585
\bibitem[Mahmoud et al.(2019)]{mah19} Mahmoud, R.~D., Done, C., \& De Marco, B.\ 2019, \mnras, 486, 2137  
\bibitem[Miller et al.(2006a)]{mil06} Miller, J.~M., Raymond, J., Fabian, A.~C., et al.\ 2006a, \nat, 441, 953
\bibitem[Miller et al.(2006b)]{mil06b} Miller, J. M., Raymond, J., Homan, J., et al.\ 2006b, \apj, 646, 394
\bibitem[Miller et al.(2008)]{mil08} Miller, J. M., Raymond, J., Reynolds, C.~S., et al.\ 2008, \apj, 680, 1359
\bibitem[Miller et al.(2012)]{mil12} Miller, J.~M., Raymond, J., Fabian, A.~C., et al.0\ 2012, \apjl, 759, 6 
\bibitem[Mitsuda et al.(1984)]{mit84} 
Mitsuda, K., Inoue, H., Koyama, K., et al.\ 1984, \pasj, 36, 741
\bibitem[Neilsen et al.(2009)]{nei09} Neilsen, J., \& Lee, J.~C.\ 2019, \nat, 458, 481
\bibitem[Neilsen et al.(2016)]{nei16} Neilsen, J., Rahoui, F., Homan, J., 
\& Buxton, M.\ 2016, \apj, 822, 20
\bibitem[Nomura et al.(2016)]{nom16} Nomura, M., Ohsuga, K., Takahashi, H.~R., Wada, K., \& Yoshida, T.\ 2016, \pasj, 68, 16
\bibitem[Ostriker et al.(1991)]{ost91} Ostriker, E.~C., McKee, C.~F., \& Klein, R.~I.\ 1991, \apj, 377, 593
\bibitem[Proga et al.(2000)]{pro00} Proga, D., Stone, J.~M., \& Kallman, T.~R.\ 2000, \apj, 543, 686
\bibitem[Proga \& Kallman(2002)]{pro02} Proga, D., \& Kallman, T.~R.\ 2002, \apj, 565, 455 
\bibitem[Ponti et al.(2012)]{pon12} Ponti, G., Fender, R.~P., Begelman, M.~C., e
t al.\ 2012, \mnras, 422, L11
\bibitem[Shidatsu et al.(2014)]{shi14} Shidatsu, M., Ueda, Y., \& Yamada, S., et al.\ 2014, \apj, 789, 100
\bibitem[Shidatsu et al.(2016)]{shi16} Shidatsu, M., Done, C., \& Ueda, Y.\ 2016, \apj, 823, 15
\bibitem[Steiner et al.(2009)]{ste09} Steiner,~J.~F., Narayan,~R., McClintock,~J.~E.,
\& Ebisawa,~K.\ 2009, \pasp, 121,1279
\bibitem[Steiner et al.(2012)]{ste12} Steiner, J.~F. McClintock, J.~E., \&  Reid, M.~J.\ 2012, \apj, 745, 7
\bibitem[Tomaru et al.(2018)]{tom18} Tomaru, R., Done, C., Odaka, H., Watanabe, S., \& Takahashi, T.\ 2018, \mnras, 476, 1776
\bibitem[Tomaru et al.(2019)]{tom19} Tomaru, R., Done, C., Ohsuga, K., Nomura, M., \& Takahashi, T.\ 2019, \mnras, submitted
\bibitem[Ueda et al.(2001)]{ued01} Ueda, Y., Asai, K., Yamaoka, K., Dotani, T., \& Inoue, H.\ 2001 \apj, 556, L87
\bibitem[Ueda et al.(2009)]{ued09} Ueda, Y., Yamaoka, K., \& Remillard, R.\ 2009 \apj, 695, 888
\bibitem[Ueda et al.(2010)]{ued10} Ueda, Y., Honda, K., Takahashi, H., et al.\ 2010, \apj, 713, 257
\bibitem[Uttley \& Klein-Wolt (2015)]{utt15} Uttley P. \& Klein-Wolt, M., 
2015, \mnras, 451, 475
\bibitem[Waters, \& Proga(2018)]{wat18} Waters, T., \& Proga, D.\ 2018, \mnras, 481, 2628
\bibitem[Wilms et al.(2000)]{wil00} Wilms, J., Allen, A., \& McCray, R.\ 2000, \apj, 542, 914
\bibitem[Woods et al.(1996)]{woo96} Woods, D.~T., Klein, R.~I., Castor, J.~I., McKee, C.~F., \& Bell, 
J.~B.\ 1996, \apj, 461, 767
\bibitem[Yamada et al.(2013)]{yam13} Yamada, S., Makishima, K., Done, S., et al.\ 2013, \pasj, 65, 80

\end{thebibliography}
\end{document}